\documentclass[10pt,a4paper]{article}
\usepackage[utf8]{inputenc}
\usepackage[english]{babel}

\usepackage{amsmath}
\usepackage{amsfonts}
\usepackage{amssymb}
\usepackage{amsthm}
\usepackage{mathrsfs}

\usepackage[colorlinks,citecolor=blue,urlcolor=blue,linkcolor=blue]{hyperref}

\usepackage[left=2cm,right=2cm,top=2cm,bottom=2cm]{geometry}

\usepackage{authblk}

\usepackage{feynmp}
\newcommand{\pd}{\partial}
\newcommand{\dd}{\mathrm{d}}
\newcommand{\ii}{\mathrm{i}}
\newcommand{\Ker}{\operatorname{Ker}}
\newcommand{\Img}{\operatorname{Im}}

\newcommand{\OO}{\mathcal{O}}

\title{How Far Can Vierbeins Simplify Gravity?}
\author[1,2]{Boris Latosh \thanks{latosh@theor.jinr.ru}}
\affil[1]{Bogoliubov Laboratory of Theoretical Physics, JINR, Dubna, 141980, Russia}
\affil[2]{Dubna State University, Universitetskaya str. 19, Dubna, 141982, Russia}
\date{}

\begin{document}

\maketitle

\begin{abstract}
  We construct a classical formulation of gravity in terms of the vierbein that makes the Hilbert action polynomial in gravitational perturbations and identify some matter models that admit a polynomial coupling to such perturbations as well. We introduce the density vierbein variables that factorize the inverse metric density while retaining an explicit local Lorentz frame. With the introduction of a new auxiliary field, one can construct an action that is polynomial in such variables and their perturbations. Constructed the BRST complex for the model and derived propagators and interaction rules. In four dimensions, a scalar field with the canonical kinetic term and a potential admits a finite number of coupling terms with the density vierbein perturbations. Among all Horndeski gravity models, only a narrow subclass admits a polynomial coupling to the density vierbein perturbations. For instance, the Einstein--scalar--Gauss--Bonnet model always has an infinite tower of interactions with the density vierbein perturbation. Similarly, the standard kinetic term for the Dirac fermions and the standard kinetic term for a vector field both admit an infinite tower of interactions. Consequently, the density vierbein variables provide a great simplification for scalar-gravitational couplings, but the simplification is not strong enough to produce a realistic model with a finite number of matter-gravity terms.
\end{abstract}

\section{Introduction}

Perturbative quantum gravity expands the spacetime metric around a classical background,
\begin{align}\label{eq:the_perturbative_expansion}
  g_{\mu\nu}=\overline g_{\mu\nu}+\kappa h_{\mu\nu}.
\end{align}
For the Hilbert action the perturbative metric \eqref{eq:the_perturbative_expansion} produces the effective field theory of gravity \cite{Feynman:1963ax,DeWitt:1967yk,DeWitt:1967ub,DeWitt:1967uc,tHooft:1974toh,Donoghue:1994dn,Burgess:2003jk}. Although~\eqref{eq:the_perturbative_expansion} is linear in $h_{\mu\nu}$, the inverse metric and volume density are infinite Taylor series in $\kappa$. The resulting action therefore contains graviton interactions of arbitrarily high valence, so it becomes a theory with an infinite tower of interactions.

These interactions make direct Feynman-diagram calculations demanding. At fixed perturbative order and fixed external multiplicity, only finitely many diagrams contribute, but their tensor expressions can be lengthy \cite{DeWitt:1967uc,Sannan:1986tz}. Gravity-matter rules of arbitrary valence and their automated implementations address this difficulty \cite{Prinz:2020bkp,Latosh:2022ydd,Latosh:2023zsi,Latosh:2024lhl,Latosh:2025vax}. In particular, package \texttt{FeynGrav} that operates within the \texttt{FeynCalc} ecosystem~\cite{Mertig:1990an,Shtabovenko:2016sxi,Shtabovenko:2020gxv,Shtabovenko:2023idz}. Automation organizes the expansion, while a change of gravitational variables can alter the number of primitive interaction vertices themselves.

Cheung and Remmen developed reorganizations of the pure gravity action~\cite{Cheung:2016say} and showed that an auxiliary-field formulation based on the inverse metric density $\sqrt{-g}g^{\mu\nu}$ makes the Hilbert action purely cubic~\cite{Cheung:2017kzx}. Their construction establishes that the infinite tower in the conventional metric expansion is compatible with a finite set of primitive interactions in other variables. Extending this description to a local Lorentz frame allows the gravitational and spinor sectors to be expressed in the same variables since the standard covariant derivative of a spinor requires a vierbein and spin connection~\cite{Fock:1929vt,Birrell:1982ix,Mei:2007xk,Parker:2009uva}. It is important to establish whether these new variables allow both the gravitational and matter sectors of the theory (especially fermions) to admit a finite number of interaction terms.

In this work, we present the density vierbein variables and analyze their coupling to matter. We resolve the algebraic degeneracy of the Hilbert quadratic form, construct its constrained auxiliary representation, and give the gauge-fixing and perturbative structures associated with both diffeomorphisms and local Lorentz transformations. We then determine the class of Horndeski gravity models that admit a finite number of couplings to the density vierbein variables. We examine the standard fermion and vector actions and find that they still admit an infinite tower of interactions. These results connect the polynomial formulation of general relativity to the conditions under which the polynomial interaction structure survives the addition of matter.

We introduce the density vierbein
\begin{align}
  (E^\mu)_a &\overset{\text{def}}{=}\sqrt e\,(e^\mu)_a, &
  e &= \det(e_\mu)^a=\sqrt{-g}>0,
\end{align}
which factorizes the inverse metric density:
\begin{align}
  \mathfrak g^{\mu\nu}\overset{\text{def}}{=}\sqrt{-g}g^{\mu\nu} =\eta^{ab}(E^\mu)_a(E^\nu)_b.
\end{align}
Here $(e^\mu)_a$ is the inverse of the standard vierbein. The density weight of $E$ is $1/2$, and its inverse and its connection must be distinguished from their ordinary vierbein counterparts.

Modulo a boundary term, the Hilbert action becomes schematically
\begin{align}
  S_{\rm H}=\frac12\int\dd^dx\,\partial E\cdot\OO[E]\cdot\partial E,
\end{align}
where $\partial E$ denotes a density frame derivative and $\OO[E]$ is algebraic and polynomial in the density vierbein. We identify the $\OO$ kernel and choose a complement on which the map is invertible. The resulting inverse map lets us construct a constrained auxiliary-field action with a finite perturbative expansion. This algebraic kernel analysis is separate from gauge fixing. The auxiliary image constraint remains field-dependent and must be preserved beyond the free theory.

The local Lorentz frame introduces a gauge symmetry in addition to diffeomorphisms. We construct the BRST transformations for the density vierbein perturbations that fix both symmetries and derive the flat-background propagators and interaction coefficients. The constrained gravitational polynomial has six cubic and quartic interaction classes. The chosen linear gauge functions give four classes of interaction with the ghosts, and all such interactions are cubic. The density vierbein propagator includes an algebraic antisymmetric contribution as well as the pole term.

Coupling to matter requires a separate treatment. First, derive the standard Dirac action in density vierbein variables and show that it has an infinite tower of interaction terms. We then determine the finite class of Horndeski gravity that admits a finite number of interaction terms with gravity. The class contains the scalar potentials, canonical scalar kinetic terms, and a Gauss--Bonnet-like coupling $f(\phi) \, R$. We separately examine the Einstein--scalar--Gauss--Bonnet gravity and find that it has an infinite tower of interaction terms with gravity. Lastly, we examine the canonical kinetic term for a vector field and find that it also admits an infinite tower of interactions with gravity. This implies that the minimally coupled Standard Model also has an infinite tower of interactions with gravity despite the simplification of the Hilbert sector.

The paper is organized as follows. Section~\ref{sec:Hilbert_action} introduces the density vierbein and implements it in the Hilbert action. Section~\ref{sec:Hilbert_action_perturbation} constructs the auxiliary field formulation. Section~\ref{sec:BRST} develops the BRST transformations, introduces the gauge-fixing term, and derives perturbative coefficients. Section~\ref{sec:Dirac_action} examines Dirac fermions, while Section~\ref{sec:matter_couplings} examines coupling to scalars and vectors. Section~\ref{sec:conclusions} discusses the scope and implications of the results. Appendices~\ref{app:vierbein-conventions} and~\ref{app:density-spin connection} collect geometric conventions and identities. Appendix~\ref{app:feynman-rules} gives the background propagators and explicit interaction coefficients of the constrained polynomial.

\section{Density vierbein variables}\label{sec:Hilbert_action}

This section fixes the conventions used throughout the paper and introduces the density vierbein variables. We assume $d>2$ and work locally with an invertible vierbein of positive determinant. We recall only the necessary part of the ordinary vierbein formalism below.

Let $(e_\mu)^a$ be a vierbein, $(e^\mu)_a$ its inverse, and
\begin{align}
  e \overset{\text{note}}{=} \det(e_\mu)^a \not = 0
\end{align}
be the vierbein determinant. Since the vierbein determinant does not vanish, the vierbein $(e_\mu)^m$ is invertible, and we choose the orientation so that
\begin{align}
  e = + \sqrt{-g}.
\end{align}
The spacetime metric is reconstructed from the vierbein by the well-known expressions
\begin{align}
  g_{\mu\nu} &= (e_\mu)^a(e_\nu)^b\eta_{ab} , & g^{\mu\nu} &= (e^\mu)_a(e^\nu)_b\eta^{ab}.
\end{align}
Greek indices are spacetime indices, while Latin indices are local Lorentz indices. Local indices are raised and lowered with $\eta_{ab}$, and spacetime indices are raised and lowered with $g_{\mu\nu}$.

The vierbein carries two kinds of indices and therefore transforms under two different groups. Under a coordinate transformation, it transforms as
\begin{align}
  (e_\mu)^a\to \mathcal{J}_\mu{}^\nu(e_\nu)^a,
\end{align}
where $\mathcal{J}_\mu{}^\nu$ is the Jacobian matrix. Under a local Lorentz transformation, it transforms as
\begin{align}
  (e_\mu)^a & \to (e_\mu)^b(J^{-1})_b{}^a, & J^a{}_cJ^b{}_d\eta_{ab} &= \eta_{cd},
\end{align}
where $J_m{}^n$ is a coordinate-dependent Lorentz matrix. This double transformation law is the reason why a perturbative theory written in terms of vierbeins has both diffeomorphism and local Lorentz gauge redundancies.

We adopt the view that a vierbein is not simply a square root of a metric. We treat it as a soldering form that relates the tangent bundle to an internal Lorentz vector bundle~\cite{Fock:1929vt,Kibble:1961ba,Hehl:1976kj,Krasnov:2020lku}. In this paper, the practical consequence is that we must keep local Lorentz covariance explicit.

The covariant derivative of a vierbein reads
\begin{align}\label{eq:ordinary_vierbein_covariant_derivative}
  \nabla_\mu(e_\nu)^a = \pd_\mu(e_\nu)^a - (\Gamma_\mu)^\rho{}_\nu(e_\rho)^a + (\omega_\mu)^a{}_b(e_\nu)^b.
\end{align}
Here $(\Gamma_\mu)^\rho{}_\nu$ is the affine connection and $(\omega_\mu)^a{}_b$ is the spin connection. We use notation $(\Gamma_\mu)^\rho{}_\nu$ because it explicitly shows which index of the connection corresponds to the index of the covariant derivative. The spin connection appears in the covariant derivative of Dirac spinors, so that the corresponding convention will be important in Section~\ref{sec:Dirac_action}.

In general relativity, the affine connection becomes the Levi-Civita connection for the metric $g_{\mu\nu}$. Equivalently, it is the unique affine connection satisfying metric compatibility and the torsion-free condition,
\begin{align}\label{eq:metric_compatibility_and_torsion_free}
  \nabla_\alpha g_{\mu\nu} & = 0 , & T^\rho{}_{\mu\nu} &=0 .
\end{align}
Consequently, coefficients $(\Gamma_\mu)^{\rho}{}_\nu$ become the Christoffel symbols $\Gamma^\rho_{\mu\nu}$.

In the vierbein formulation, one must enforce compatibility between spacetime parallel transport and local Lorentz parallel transport with the tetrad postulate
\begin{align}\label{eq:tetrad_postulate}
  \nabla_\mu(e_\nu)^a=0.
\end{align}
Together with local Lorentz metric compatibility,
\begin{align}\label{eq:local_lorentz_metric_compatibility}
  \nabla_\mu\eta_{ab} = 0,
\end{align}
This implies spacetime metric compatibility, because
\begin{align}
  \nabla_\rho g_{\mu\nu} = \nabla_\rho\big[(e_\mu)^a(e_\nu)^b\eta_{ab}\big] =0.
\end{align}
Conversely, the tetrad postulate determines the relation between the spin connection, affine connection, and vierbeins
\begin{align}\label{eq:ordinary_spin_connection_from_tetrad_postulate}
  (\omega_\mu)^a{}_b = (\Gamma_\mu)^\rho{}_\nu(e_\rho)^a(e^\nu)_b - \pd_\mu(e_\nu)^a(e^\nu)_b.
\end{align}

One can write the torsion-free condition in a local Lorentz form with Cartan's first structure equation
\begin{align}\label{eq:cartan_first_structure_equation}
  T^a = \dd e^a+\omega^a{}_b\wedge e^b=0.
\end{align}
This condition selects the ordinary Levi-Civita spin connection and gives the usual second-order vierbein formulation of the Hilbert action. In particular, it ensures that the ordinary spin connection is antisymmetric
\begin{align}
  (\omega_\mu)_{ab}=-(\omega_\mu)_{ba}.
\end{align}

We introduce the density vierbein variables as follows. Since the density vierbein variables shall factorize $\sqrt{-g} g^{\mu\nu}$ and the ordinary inverse vierbein factorizes $g^{\mu\nu} = (e^\mu)_m (e^\nu)_n \eta^{mn}$, the required variables should only be equipped with a proper power of the vierbein determinant.

The definition of the density vierbein variables reads
\begin{align}
  (E^\mu)_a \overset{\text{def}}{=} \sqrt e\,(e^\mu)_a.
\end{align}
The following conditions define the inverse variable
\begin{align}
  (E_\mu)^a(E^\mu)_b &= \delta^a_b, & (E_\mu)^a(E^\nu)_a &= \delta_\mu^\nu.
\end{align}
In terms of the ordinary vierbein,
\begin{align}
  (E_\mu)^a=\frac{1}{\sqrt e}(e_\mu)^a.
\end{align}
Further, we use the following notation for the density vierbein determinant
\begin{align}
  E\overset{\text{note}}{=} \det(E^\mu)_a.
\end{align}
This determinant relates to the regular vierbein determinant
\begin{align}
  E &= e^{(d-2)/2}, & e &= E^{2/(d-2)}.
\end{align}
The central identity for the density vierbein variables reads
\begin{align}\label{eq:density_vierbein_factorisation}
  (E^\mu)_a (E^\nu)_b \eta^{ab} = \sqrt{-g}\,g^{\mu\nu}.
\end{align}
Thus, the density vierbein variables factorize the Cheung-Remmen variables and directly generalize them.

The factor $\sqrt e$ changes the diffeomorphism transformation law, so $(E^\mu)_a$ is not an ordinary vector but a contravariant spacetime vector density of weight $1/2$. Consequently, its infinitesimal transformation law reads
\begin{align}\label{eq:density_vierbein_gauge_transformation}
  \delta(E^\mu)_a = \zeta^\rho\pd_\rho(E^\mu)_a - (E^\rho)_a\pd_\rho\zeta^\mu + \cfrac12 \, (E^\mu)_a\pd_\rho\zeta^\rho + \lambda_a{}^b(E^\mu)_b,
\end{align}
where $\zeta^\mu$ is the diffeomorphism parameter and $\lambda_{ab}=-\lambda_{ba}$ is the local Lorentz parameter.

The corresponding covariant derivative of the density vierbein variables is
\begin{align}\label{eq:density_vierbein_covariant_derivative}
  \nabla_\mu(E^\alpha)_a = \pd_\mu(E^\alpha)_a + (\Gamma_\mu)^\alpha{}_\beta(E^\beta)_a - (E^\alpha)_b(\Omega_\mu)^b{}_a
\end{align}
with $(\Omega_\mu)^a{}_b$ denoting the corresponding connection which we call {\it density spin connection}. There is no explicit density weight trace term in~\eqref{eq:density_vierbein_covariant_derivative}, because that contribution is absorbed into $(\Omega_\mu)^a{}_b$. Its relation to the ordinary spin connection reads
\begin{align}\label{eq:density_spin_connection_definition}
  (\Omega_\mu)^a{}_b \overset{\text{def}}{=} (\omega_\mu)^a{}_b + \cfrac{1}{d-2} ~  (E_\sigma)^s\pd_\mu(E^\sigma)_s\,\delta^a_b  = (\omega_\mu)^a{}_b + \pd_\mu\ln\sqrt e\,\delta^a_b.
\end{align}
The trace term absorbs the contribution coming from the density weight of $(E^\mu)_a$. The ordinary spin connection is antisymmetric, but the density spin connection has a symmetric part due to the same trace term
\begin{align}\label{eq:density_spin_connection_compatibility}
  (\Omega_\mu)_{ab} + (\Omega_\mu)_{ba} = \frac{2}{d-2} (E_\sigma)^s\pd_\mu(E^\sigma)_s\,\eta_{ab}.
\end{align}

It is useful to express the density spin connection explicitly in terms of the density vierbein variables. Local Lorentz indices on $E$ are raised and lowered with $\eta_{ab}$. With the density frame component
\begin{align}
  (\Omega_m)_{ab} \overset{\text{def}}{=} (E^\mu)_m(\Omega_\mu)_{ab},
\end{align}
one has
\begin{align}\label{eq:density_spin_connection_explicit}
  \begin{split}
    (\Omega_m)_{ab} = \Bigg[ & \cfrac12 \left[ (E^\sigma)_a\delta_b{}^c \!-\! (E^\sigma)_b\delta_a{}^c \right] (E_\alpha)_m \!+\! \cfrac12 \left[ (E^\sigma)_m\delta_b{}^c \!-\! (E^\sigma)_b\delta_m{}^c \right] (E_\alpha)_a \\
      & -\! \cfrac12 \left[ (E^\sigma)_m\delta_a{}^c \!-\! (E^\sigma)_a\delta_m{}^c \right] (E_\alpha)_b \!+\! \cfrac{1}{d-2} \left[ (E^\sigma)_m\eta_{ab} \!+\! (E^\sigma)_b\eta_{ma} \!-\! (E^\sigma)_a\eta_{mb} \right] (E_\alpha)^c \Bigg] \pd_\sigma(E^\alpha)_c .
  \end{split}
\end{align}
This formula shows two facts that will be important below. First, the density spin connection $\Omega$ is linear in the first derivatives of the density vierbein variables. Second, the coefficients multiplying these derivatives contain inverse density vierbein variables.

The Hilbert action in $d$ dimensions is defined by
\begin{align}\label{eq:hilbert_action_definition}
  S_{\rm H} \overset{\text{def}}{=}  -\cfrac{2}{\kappa^{d-2}} ~ \int\dd^d x ~\sqrt{-g}\,R. 
\end{align}
Here $\kappa$ is the gravitational coupling with the mass dimension $-1$. The trace shift in \eqref{eq:density_spin_connection_definition} does not modify the curvature contribution to the Hilbert action since its field strength is proportional to $\pd_\mu\pd_\nu\ln\sqrt e-\pd_\nu\pd_\mu\ln\sqrt e$, which vanishes. In terms of the density vierbein variables and the density spin connection, the Hilbert action reads
\begin{align}\label{eq:hilbert_action_density_spin_connection}
  S_\text{H} = -\cfrac{2}{\kappa^{d-2}} ~ \int \dd^d x ~ \eta^{mn} \, (E^\mu)_s \, (E^\nu)_n \Big[ \pd_\mu(\Omega_\nu)^s{}_m - \pd_\nu(\Omega_\mu)^s{}_m + (\Omega_\mu)^s{}_c(\Omega_\nu)^c{}_m - (\Omega_\nu)^s{}_c(\Omega_\mu)^c{}_m \Big].
\end{align}

To simplify the action further, we integrate by parts terms with $\partial \Omega$ and separate the boundary term. After this, we use expression \eqref{eq:density_spin_connection_explicit} for the density spin connection and introduce the density frame derivative
\begin{align}\label{eq:density_frame_derivative}
  \pd_s(E^\alpha)_a \overset{\text{note}}{=} (E^\sigma)_s\pd_\sigma(E^\alpha)_a.
\end{align}
The symbol $\pd_s$ denotes this density frame component of the derivative, and {\it it is not} a derivative with respect to a coordinate $x^s$.

The resulting form of the Hilbert action is
\begin{align}\label{eq:hilbert_action_density_vierbein_operator}
  \begin{split}
    S_\text{H} &= \int \dd^d x ~ \cfrac12 ~ \pd_{s_1}(E^{\alpha_1})_{a_1} \, \OO^{s_1}{}_{\alpha_1}{}^{a_1}{}^{s_2}{}_{\alpha_2}{}^{a_2} \, \pd_{s_2}(E^{\alpha_2})_{a_2} + S_\text{boundary}, \\
    S_\text{boundary} &= -\cfrac{2}{\kappa^{d-2}} \int_{\partial M}\dd^{d-1}S_\mu\, \left[ (E^\mu)_a(E^\nu)_b - (E^\mu)_b(E^\nu)_a \right] (\Omega_\nu)^{ab}.
  \end{split}
\end{align}
The bulk action \eqref{eq:hilbert_action_density_vierbein_operator} is a quadratic form with respect to $\partial_s (E^\alpha)_a$, so we will call it the Hilbert form. In turn, the operator~$\OO$ reads
\begin{align}\label{eq:density_vierbein_operator}
  \begin{split}
    \OO^{s_1}{}_{\alpha_1}{}^{a_1}{}^{s_2}{}_{\alpha_2}{}^{a_2} = \cfrac{2}{\kappa^{d-2}} \Bigg[ &  + \eta^{a_1a_2} (E_{\alpha_2})^{s_1}(E_{\alpha_1})^{s_2} - 2\eta^{a_1a_2} (E_{\alpha_1})^{s_1}(E_{\alpha_2})^{s_2} \\
      & -\eta^{s_1a_2}\eta^{s_2a_1} (E_{\alpha_1})_m(E_{\alpha_2})^m - \eta^{s_2a_1}(E_{\alpha_1})^{a_2}(E_{\alpha_2})^{s_1} -\eta^{s_1a_2}(E_{\alpha_2})^{a_1}(E_{\alpha_1})^{s_2} \\
      & + \eta^{a_1a_2}\eta^{s_1s_2} (E_{\alpha_1})_m(E_{\alpha_2})^m + \eta^{s_1s_2}(E_{\alpha_1})^{a_2}(E_{\alpha_2})^{a_1} \!-\! \cfrac{2}{d-2} \, \eta^{s_1s_2} (E_{\alpha_1})^{a_1}(E_{\alpha_2})^{a_2} \Bigg].
  \end{split}
\end{align}

Action~\eqref{eq:hilbert_action_density_vierbein_operator} is the key result of this section. The Hilbert action is quadratic in the variables $\pd_s(E^\alpha)_a$, while all nonlinearity is contained algebraically in the operator $\OO[E]$. The next step is to turn this form into a finite perturbative action. This cannot be done by inverting $\OO$ on the full space of variables $\pd_s(E^\alpha)_a$, because $\OO$ has a nontrivial kernel. This is an algebraic degeneracy of the derivative quadratic form, and it must be distinguished from the differential gauge identities of the action. In the next section, we identify the kernel, choose a complement, and introduce the auxiliary field.

\section{Auxiliary field formulation}\label{sec:Hilbert_action_perturbation}

In Section~\ref{sec:Hilbert_action}, we brought the Hilbert action to the form \eqref{eq:hilbert_action_density_vierbein_operator}. The operator $\OO[E]$ is algebraic in density vierbein variables, while derivatives enter only through the density frame derivatives $\pd_s(E^\alpha)_a$. We first establish classical equivalence to a formulation with one auxiliary field subjected to certain algebraic restrictions. This is sufficient to cast the action into polynomial form. This construction follows the same general strategy as the Cheung-Remmen formulation of pure gravity. One exchanges a part of the nonlinear structure for an auxiliary field whose equation of motion reconstructs the original second-order action~\cite{Cheung:2016say,Cheung:2017kzx}.

It may appear that one needs only to use the density spin connection $\Omega[E]$ as the auxiliary variable, since this approach has a striking analogy with the Palatini formulation. However, this approach is misleading. The density spin connection $\Omega[E]$ is subject to the compatibility condition~\eqref{eq:density_spin_connection_compatibility}. Let us see what will happen when we introduce perturbations to the density vierbein variables
\begin{align}
  (E^\mu)_a = \delta^\mu_a + \kappa \, (\Theta^\mu)_a,
\end{align}
The inverse density vierbein $(E_\mu)^a$ is an infinite series in $\kappa$. Consequently, the trace term $(E_\sigma)^s\pd_\mu(E^\sigma)_s$ in~\eqref{eq:density_spin_connection_compatibility} also contains arbitrarily high powers of $\kappa$. This leaves the geometric connection $\Omega[E]$ no choice but to be an infinite series in the perturbation, which will spawn an infinite tower of interaction terms. Therefore, one must introduce the auxiliary field that is specifically adapted to the Hilbert quadratic form.

To proceed and identify the correct auxiliary field, we shall rewrite the operator $\OO$ in all-low local-frame notation. To keep the notation compact, we define
\begin{align}\label{eq:X_definition_section3}
  X_{ijk} \overset{\text{note}}{=} (E_\alpha)_j\pd_i(E^\alpha)_k.
\end{align}
In this notation, the Hilbert action takes the following form
\begin{align}\label{eq:hilbert_action_X_notation}
  S_\text{H} = \frac12\int\dd^d x\,X_{i_1j_1k_1}\OO^{i_1j_1k_1i_2j_2k_2}X_{i_2j_2k_2},
\end{align}
where we omitted the boundary term. The action of $\OO$ on an arbitrary $X_{ijk}$ reads
\begin{align}\label{eq:operator_O_all_low_section3}
  (\OO X)_{ijk} = \cfrac{2}{\kappa^{d-2}} \Big[ -X_{jki} +X_{ikj} -X_{kji} -X_{kij} +X_{ijk} +X_{jik} - 2\, \eta_{ij}X^s{}_{sk} -\cfrac{2}{d-2} \, \eta_{jk}X_{is}{}^s \Big].
\end{align}

The operator $\OO$ has a nontrivial kernel spanned by the following objects:
\begin{align}\label{eq:kernel_O_section3}
  X_{ijk} &= S_{ijk}+S_{jki}-S_{ikj} + \eta_{ik}S^s{}_{sj} - \eta_{jk}S^s{}_{si}, & S_{ijk} &=S_{jik}.
\end{align}
This degeneracy is purely algebraic. Consequently, one cannot invert $\OO$ on the full tensor space.

To select a complement to $\Ker\OO$, it is useful to define the kernel projector
\begin{align}\label{eq:kernel_projector_section3}
  (\mathcal{P} X)_{abc} \overset{\text{def}}{=} \frac12\left(X_{abc}+X_{bac}\right) + \frac12\left(X_{bca}+X_{cba}\right) - \frac12\left(X_{acb}+X_{cab}\right) + \eta_{ac}X^s{}_{sb} - \eta_{bc}X^s{}_{sa}.
\end{align}
Direct calculation shows that $\mathcal{P}$ is a projector since $\mathcal{P}^2 = \mathcal{P}$ and it projects on $\Ker \OO$ since $\mathcal\OO\mathcal{P} X = 0$. Consequently, the projector on the complement of $\Ker \OO$ reads
\begin{align}\label{eq:X_perp_definition_section3}
  X^\perp \overset{\text{def}}{=} \left( 1 - \mathcal{P} \right) X
\end{align}
and the following identities hold
\begin{align}\label{eq:operator_kernel_projector_identities_section3}
  \OO X &= \OO X^\perp, &  X\OO X & =X^\perp\OO X^\perp.
\end{align}
The chosen complement is characterized by $X^\perp_{ijk}+X^\perp_{jik}=0$.

The inverse tensor $\OO^{-1}$ is therefore defined only on the complement of $\ker\OO$. For $A\in\Img\OO$ we require
\begin{align}\label{eq:O_inverse_on_complement_section3}
  \OO\OO^{-1}A &= A, & \OO^{-1}\OO X &= X^\perp.
\end{align}
This is the inverse tensor that we will use for the auxiliary field construction.

We now introduce the auxiliary field $A^i{}_{\alpha}{}^a$, constrained to the corresponding nondegenerate space (the compliment of $\ker\OO$). The auxiliary field action reads
\begin{align}\label{eq:auxiliary_field_action_section3}
  S_\text{H}[E,A] = \int\dd^d x \left[ -\cfrac12\, A^{i_1}{}_{\alpha_1}{}^{a_1} (\OO^{-1})_{i_1}{}^{\alpha_1}{}_{a_1}{}_{i_2}{}^{\alpha_2}{}_{a_2} A^{i_2}{}_{\alpha_2}{}^{a_2} + A^i{}_{\alpha}{}^a\pd_i(E^\alpha)_a \right].
\end{align}
Varying with respect to $A$, with $A$ restricted to $\Img\OO$, gives
\begin{align}\label{eq:auxiliary_field_eom_section3}
  A^i{}_{\alpha}{}^a = \OO^i{}_{\alpha}{}^a{}^j{}_{\beta}{}^b\pd_j(E^\beta)_b.
\end{align}
The right-hand side is automatically projected onto the same nondegenerate subspace. Substituting this equation back into \eqref{eq:auxiliary_field_action_section3} returns \eqref{eq:hilbert_action_density_vierbein_operator}. Thus, eliminating the constrained auxiliary field by its algebraic equation of motion recovers the Hilbert action (modulo the same boundary term). This establishes classical equivalence.

The auxiliary field $A$ is not the density spin connection. The spin connection is a geometric object fixed by metric compatibility, while $A$ is an algebraic field conjugate to $\pd_i(E^\alpha)_a$. The algebraic relation between the auxiliary field and the density spin connection is obtained by comparing the part of the Hilbert action that is linear in $\pd E$ with the coupling $A^i{}_{\alpha}{}^a\pd_i(E^\alpha)_a$. The explicit expression for the auxiliary field~$A$ in terms of density spin connection reads
\begin{align}\label{eq:auxiliary_field_density_spin_connection_relation}
  A^i{}_{\alpha}{}^a[\Omega,E] = \frac{2}{\kappa^{d-2}} \left[ (E_\alpha)^i \left( (\Omega_c)^{ac}-(\Omega_c)^{ca} \right) + (E_\alpha)^m \left( (\Omega_m)^{ia}-(\Omega_m)^{ai} \right) \right].
\end{align}
Only the antisymmetric part of the density spin connection contributes to this expression.

We now expand the density vierbein around the flat background,
\begin{align}\label{eq:density_vierbein_perturbation_section3}
  (E^\alpha)_a  = \delta^\alpha_a + \kappa \, (\theta^\alpha)_a.
\end{align}
In the flat-background expansion below, $\partial_i\overset{\text{def}}{=}\delta_i^\mu\partial_\mu$ denotes a coordinate derivative. The full density frame derivative in the unexpanded action consequently becomes
\begin{align}\label{eq:expanded_density_frame_derivative}
  (E^\sigma)_i\partial_\sigma(E^\alpha)_a =\kappa \, \partial_i(\theta^\alpha)_a+\kappa^2(\theta^\sigma)_i\partial_\sigma(\theta^\alpha)_a.
\end{align}
We introduce the shifted auxiliary field by
\begin{align}\label{eq:auxiliary_field_shift_section3}
  A^i{}_{\alpha}{}^a  = B^i{}_{\alpha}{}^a + \kappa^{3-d}\, \mathscr{P}^i{}_{\alpha}{}^a{}^j{}_{\beta}{}^b\,  \pd_j(\theta^\beta)_b.
\end{align}
With this choice, $(\theta^\alpha)_a$ has mass dimension $1$ (the canonical value in $d=4$), while $A$ and $B$ have mass dimension~$d-1$.

When one puts \eqref{eq:density_vierbein_perturbation_section3} and \eqref{eq:auxiliary_field_shift_section3} into \eqref{eq:auxiliary_field_action_section3}, the action contains a linear mixing term between $B$ and $\pd\theta$:
\begin{align}\label{eq:B_theta_mixing_section3}
  -\kappa^{3-d}\,B\,\left.\OO^{-1}\right|_0\mathscr{P}\,\pd\theta + \kappa\,B\,\pd\theta
\end{align}
Here $\left.\OO^{-1}\right|_0$ denotes $\OO^{-1}$ evaluated at the flat background. We shall choose $\mathscr P$ so it would remove the mixing term. The suitable choice is
\begin{align}\label{eq:background_P_operator_section3}
  \begin{split}
    \mathscr{P}^i{}_{\alpha}{}^a{}^j{}_{\beta}{}^b \overset{\text{def}}{=} \kappa^{d-2} \left. \OO^i{}_{\alpha}{}^a{}^j{}_{\beta}{}^b \right|_{(E^\mu)_m=\delta^\mu_m} = 2\Big[& -\eta^{ib}\delta_\beta{}^a\delta_\alpha{}^j +\eta^{ij}\delta_\alpha{}^b\delta_\beta{}^a -\eta^{ib}\eta^{ja}\eta_{\alpha\beta} -\eta^{ja}\delta_\alpha{}^b\delta_\beta{}^i \\
      & +  \eta^{ab}\eta^{ij}\eta_{\alpha\beta} -2\eta^{ab}\delta_\alpha{}^i\delta_\beta{}^j +\eta^{ab}\delta_\beta{}^i\delta_\alpha{}^j -\frac{2}{d-2}\eta^{ij}\delta_\alpha{}^a\delta_\beta{}^b \Big].
  \end{split}
\end{align}
With this definition, the quadratic part of the action is fixed
\begin{align}\label{eq:quadratic_action_identity_section3}
  &-\frac12\,B\left.\OO^{-1}\right|_0 B -\frac12\,\kappa^{2(3-d)} \pd\theta\,\mathscr{P} \left.\OO^{-1}\right|_0 \mathscr{P} \pd\theta +\kappa^{4-d}\,\pd\theta\,\mathscr{P}\,\pd\theta  = -\frac12\,\kappa^{d-2}B\,\mathscr{P}^{-1}B +\frac12\,\kappa^{4-d}\pd\theta\,\mathscr{P}\,\pd\theta .
\end{align}
The inverse tensor $\mathscr{P}^{-1}$ on the nondegenerate subspace is
\begin{align}\label{eq:background_P_inverse_section3}
  \begin{split}
    (\mathscr{P}^{-1})_{i_1}{}^{\alpha_1}{}_{a_1}{}_{i_2}{}^{\alpha_2}{}_{a_2} \overset{\text{def}}{=} \kappa^{2-d} \left. (\OO^{-1})_{i_1}{}^{\alpha_1}{}_{a_1}{}_{i_2}{}^{\alpha_2}{}_{a_2} \right|_{(E^\mu)_m=\delta^\mu_m} = \frac14\,\eta_{i_1i_2}\delta^{\alpha_2}{}_{a_1}\delta^{\alpha_1}{}_{a_2} - \frac18\,\eta_{i_1i_2}\delta^{\alpha_1}{}_{a_1}\delta^{\alpha_2}{}_{a_2}.
  \end{split}
\end{align}
The exact restriction on the shifted field is inherited from $A\in\Img\OO[E]$:
\begin{align}\label{eq:shifted_auxiliary_image_constraint}
  B+\kappa^{3-d}\mathscr P\,\partial\theta & \in\Img\OO[E], &
  \mathcal P[E]^\dagger\big(B+\kappa^{3-d}\mathscr P\,\partial\theta\big) &= 0.
\end{align}
At the flat background, $\mathscr P\partial\theta\in\Img\OO_0$, so the restriction on the free auxiliary field reduces to
\begin{align}\label{eq:no_mixing_condition_section3}
  \mathcal{P}_0^\dagger B &= 0, & \mathcal{P}_0 &\overset{\text{def}}{=}  \left. \mathcal{P}\right|_{(E^\mu)_m=\delta^\mu_m}.
\end{align}
This background condition removes the quadratic $B$-$\theta$ mixing term after the shift \eqref{eq:auxiliary_field_shift_section3}.

The inverse operator $\OO^{-1}$ has a simple, exact dependence on the density vierbein variables:
\begin{align}\label{eq:exact_O_inverse_section3}
  (\OO^{-1})_{i_1}{}^{\alpha_1}{}_{a_1}{}_{i_2}{}^{\alpha_2}{}_{a_2}[E] = \kappa^{d-2} (\mathscr{P}^{-1})_{i_1}{}^{c_1}{}_{a_1}{}_{i_2}{}^{c_2}{}_{a_2} (E^{\alpha_1})_{c_1}(E^{\alpha_2})_{c_2}.
\end{align}
Formula~\eqref{eq:density_vierbein_perturbation_section3} gives the exact finite expansion of $\OO^{-1}$:
\begin{align}\label{eq:O_inverse_expansion_section3}
  \begin{split}
    (\OO^{-1})_{i_1}{}^{\alpha_1}{}_{a_1}{}_{i_2}{}^{\alpha_2}{}_{a_2}[E] = \kappa^{d-2}\Big[ & {~} (\mathscr{P}^{-1})_{i_1}{}^{\alpha_1}{}_{a_1}{}_{i_2}{}^{\alpha_2}{}_{a_2} \\
      & + \kappa \left( \delta_\gamma{}^{\alpha_1}(\mathscr{P}^{-1})_{i_1}{}^c{}_{a_1}{}_{i_2}{}^{\alpha_2}{}_{a_2} + \delta_\gamma{}^{\alpha_2}(\mathscr{P}^{-1})_{i_1}{}^{\alpha_1}{}_{a_1}{}_{i_2}{}^c{}_{a_2} \right)(\theta^\gamma)_c \\
      & + \kappa^2 \delta_{\gamma_1}{}^{\alpha_1}\delta_{\gamma_2}{}^{\alpha_2} (\mathscr{P}^{-1})_{i_1}{}^{c_1}{}_{a_1}{}_{i_2}{}^{c_2}{}_{a_2} (\theta^{\gamma_1})_{c_1}(\theta^{\gamma_2})_{c_2} \Big].
  \end{split}
\end{align}

The exact constrained action admits a polynomial representative with six interaction classes
\begin{align}\label{eq:finite_perturbative_action_section3}
  \begin{split}
    S_\text{H}[\theta,B] \!=\! \int\!\!\dd^d x\Bigg[& -\frac12\kappa^{d-2} B^{i_1}{}_{\alpha_1}{}^{a_1} (\mathscr{P}^{-1})_{i_1}{}^{\alpha_1}{}_{a_1}{}_{i_2}{}^{\alpha_2}{}_{a_2} B^{i_2}{}_{\alpha_2}{}^{a_2} + \frac12\kappa^{4-d} \pd_{i_1}(\theta^{\alpha_1})_{a_1} \mathscr{P}^{i_1}{}_{\alpha_1}{}^{a_1}{}^{i_2}{}_{\alpha_2}{}^{a_2} \pd_{i_2}(\theta^{\alpha_2})_{a_2} \\
      &+ \kappa^{5-d}\,\mathcal{V}_{\theta\theta\theta}\,\theta\,\pd\theta\,\pd\theta - \kappa^{6-d}\,\mathcal{V}_{\theta\theta\theta\theta}\,\theta\,\theta\,\pd\theta\,\pd\theta - \kappa^{d-1}\,\mathcal{V}_{\theta BB}\,\theta\,B\,B - \kappa^{d}\,\mathcal{V}_{\theta\theta BB}\,\theta\,\theta\,B\,B \\
      &+ \kappa^{2}\,\mathcal{V}_{\theta\theta B}\,\theta\,\pd\theta \, B - \kappa^{3}\,\mathcal{V}_{\theta\theta\theta B}\,\theta\,\theta\,\pd\theta\,B \Big].
  \end{split}
\end{align}
We suppress indices on the interaction terms. The symbols $\mathcal V$ denote tensors built from $\eta$, Kronecker deltas, $\mathscr{P}$, and $\mathscr{P}^{-1}$. The expressions for $\mathcal{V}$ are not needed for the structural argument, and we present them in Appendix~\ref{app:feynman-rules}.

This construction does not fix the gauge symmetry. The kernel analysis above is only the algebraic step that makes the auxiliary field well defined on the nondegenerate part of the Hilbert quadratic form. The density vierbein variables still carry diffeomorphism and local Lorentz gauge redundancies. Thus the kernel analysis and the auxiliary field construction are independent of the later choice of gauge fixing. The next section constructs the corresponding BRST transformations, chooses gauge-fixing functions, and derives the gauge-fixed perturbative action.

\section{Gauge Fixing and Perturbative Coefficients}\label{sec:BRST}

The auxiliary field formulation constructed in Section~\ref{sec:Hilbert_action_perturbation} gives a finite representation of the Hilbert action, but it still contains gauge redundancy. In this section, we develop the gauge and BRST transformations, the gauge fixing, and the background propagators and interaction coefficients of the constrained polynomial. For generality, we first work over an arbitrary background density vierbein. We evaluate the explicit quadratic kernels and interaction coefficients about the flat background.

We split the density vierbein variables as
\begin{align}\label{eq:background_split_section4}
  (E^\mu)_m=(\overline E^\mu)_m + \kappa(\theta^\mu)_m.
\end{align}
We hold the background field $(\overline E^\mu)_m$ fixed. It defines the background metric $\overline g$ and background connections. We first give the gauge and BRST transformations of the density vierbein perturbations on this arbitrary background, then specialize the auxiliary-field shift and perturbative coefficients to the flat background of Section~\ref{sec:Hilbert_action_perturbation}. An extension of the gauge transformations to the shifted auxiliary field $B$ must preserve the constrained action and~\eqref{eq:shifted_auxiliary_image_constraint}, taking account of the defining shift. We do not specify a BRST realization on independent auxiliary variables here. The gauge-fixing functions below depend only on $\theta$, so their BRST variations are fully determined by the displayed transformations of $\theta$ and the gauge-fixing fields.

The perturbative field content is
\begin{align}
  (\theta^\mu)_m, \qquad B^i{}_{\alpha}{}^a,
\end{align}
and the ghost, antighost, and Nakanishi-Lautrup fields\footnote{This terminology is used for the standard BRST gauge-fixing auxiliary fields and should not be confused with the auxiliary fields $A$ and $B$.} are
\begin{align}
  c^\mu, \qquad \chi_m{}^n, \qquad \overline c_\mu, \qquad \overline\chi_m{}^n, \qquad b_\mu, \qquad \mathfrak{b}_m{}^n.
\end{align}
The fields $c^\mu$ and $\chi_m{}^n$ are Grassmann-odd ghosts for diffeomorphisms and local Lorentz rotations. The antighosts are also Grassmann-odd, whereas the Nakanishi-Lautrup fields are Grassmann-even. The local Lorentz fields are antisymmetric,
\begin{align}
  \chi_{mn} &=-\chi_{nm}, & \overline\chi_{mn} &=-\overline\chi_{nm}, & \mathfrak{b}_{mn} &=-\mathfrak{b}_{nm}.
\end{align}

For a diffeomorphism parameter $\zeta^\mu$ and an antisymmetric local Lorentz parameter $\lambda_{mn}$, the fixed-background split gives
\begin{align}\label{eq:gauge_transformation_theta_section4}
  \begin{split}
    \delta(\theta^\mu)_m={}&\frac1\kappa\Big[\zeta^\rho\partial_\rho(\overline E^\mu)_m-(\overline E^\rho)_m\partial_\rho\zeta^\mu+\frac12(\overline E^\mu)_m\partial_\rho\zeta^\rho+\lambda_m{}^n(\overline E^\mu)_n\Big]\\
    &+\zeta^\rho\partial_\rho(\theta^\mu)_m-(\theta^\rho)_m\partial_\rho\zeta^\mu +\frac12(\theta^\mu)_m\partial_\rho\zeta^\rho+\lambda_m{}^n(\theta^\mu)_n.
  \end{split}
\end{align}
We use the coordinate ghost basis so $c^\mu$ and $\chi_m{}^n$ correspond directly to $\zeta^\mu$ and $\lambda_m{}^n$. The BRST differential is Grassmann odd, obeys the graded Leibniz rule, commutes with coordinate derivatives, and leaves the background fixed. Its transformations are
\begin{align}\label{eq:brst_transformations_background_section4}
  \begin{split}
    \delta_\text{B}(\theta^\mu)_m={}&\frac1\kappa\Big[c^\rho\partial_\rho(\overline E^\mu)_m-(\overline E^\rho)_m\partial_\rho c^\mu+\frac12(\overline E^\mu)_m\partial_\rho c^\rho+\chi_m{}^n(\overline E^\mu)_n\Big]\\
    &+c^\rho\partial_\rho(\theta^\mu)_m-(\theta^\rho)_m\partial_\rho c^\mu +\frac12(\theta^\mu)_m\partial_\rho c^\rho+\chi_m{}^n(\theta^\mu)_n,\\
    \delta_\text{B}c^\mu={}&c^\rho\partial_\rho c^\mu,\\
    \delta_\text{B}\chi_m{}^n={}&c^\rho\partial_\rho\chi_m{}^n+\chi_m{}^s\chi_s{}^n,\\
    \delta_\text{B}\overline c_\mu={}&b_\mu, \\
    \delta_\text{B}b_\mu =& 0,\\
    \delta_\text{B}\overline\chi_{mn}=&\mathfrak b_{mn},\\
    \delta_\text{B}\mathfrak b_{mn}=&0.
  \end{split}
\end{align}
The displayed transformations of the density vierbein perturbations, ghosts, antighosts, and Nakanishi--Lautrup fields are nilpotent. Nilpotency follows from the diffeomorphism and local Lorentz algebra with the graded Leibniz rule. In this coordinate basis, there is no explicit background-curvature term in $\delta_\text{B}\chi$.

Let $\mathcal{G}_\mu[\theta;\overline E]$ and $\mathcal{Z}_{mn}[\theta;\overline E]$ be background-covariant gauge-fixing functions for diffeomorphisms and local Lorentz transformations. For generality, we also introduce background tensors $Y^{\mu\nu}$ and $Z^{mnrs}$, which can be differential operators, and which are BRST invariant, formally self-adjoint with respect to the background measure, and invertible on the relevant gauge-fixing subspaces. In these expressions, the gauge functions, ghosts, antighosts, and Nakanishi--Lautrup fields have density weight zero. The gauge-fixing form reads
\begin{align}\label{eq:gauge_fixing_fermion_general_section4}
  \Psi \overset{\text{def}}{=} \kappa^{4-d}\int\dd^dx\sqrt{-\overline g}\, \left[ \overline c_\mu Y^{\mu\nu} \left( \mathcal{G}_\nu-\frac{\epsilon}{2}\kappa^2b_\nu \right) + \overline\chi_{mn}Z^{mnrs} \left( \mathcal{Z}_{rs}-\frac{\rho}{2}\kappa^2\mathfrak{b}_{rs} \right) \right].
\end{align}
Here $\epsilon$ and $\rho$ are gauge parameters. The gauge-fixing and ghost action is BRST exact:
\begin{align}\label{eq:gf_gh_brs_exact_section4}
  S_\text{gf+gh}=\delta_\text{B}\Psi,
\end{align}
so this gauge-fixing sector is BRST invariant before eliminating the Nakanishi--Lautrup fields. Eliminating those fields by their algebraic equations gives the reduced action
\begin{align}\label{eq:reduced_gf_gh_action_general_section4}
  S_\text{gf+gh}^\text{red} = \kappa^{4-d}\int\dd^dx\sqrt{-\overline g}\, \Big[& \frac{1}{2\epsilon\kappa^2} \, \mathcal{G}_\mu Y^{\mu\nu}\mathcal{G}_\nu - \overline c_\mu Y^{\mu\nu}\delta_\text{B}\mathcal{G}_\nu + \frac{1}{2\rho\kappa^2}\mathcal{Z}_{mn}Z^{mnrs}\mathcal{Z}_{rs} - \overline\chi_{mn}Z^{mnrs}\delta_\text{B}\mathcal{Z}_{rs} \Big].
\end{align}
This formula determines the flat-background gauge-fixing and ghost coefficients.

Further we proceed with the flat background $(\overline E^\mu)_a = \delta^\mu_a$, $\overline\Gamma=0$, $\overline\omega=0$, and $\overline R=0$. All explicit propagators and interaction coefficients below refer to this flat-background expansion. The BRST transformations reduce to
\begin{align}\label{eq:brst_theta_flat_section4}
  \begin{split}
    \delta_\text{B}(\theta^\mu)_m =& -\frac{1}{\kappa}\partial_m c^\mu + \frac{1}{2\kappa}\delta_m^\mu\partial_\rho c^\rho + \frac{1}{\kappa}\chi_m{}^n\delta_n^\mu + c^\rho\partial_\rho(\theta^\mu)_m - (\theta^\rho)_m\partial_\rho c^\mu + \frac12(\theta^\mu)_m\partial_\rho c^\rho + \chi_m{}^n(\theta^\mu)_n, \\
    \delta_\text{B}c^\mu =&\ c^\rho\partial_\rho c^\mu, \\
    \delta_\text{B}\chi_m{}^n =&\ c^\rho\partial_\rho\chi_m{}^n+\chi_m{}^s\chi_s{}^n.
  \end{split}
\end{align}

We choose gauge-fixing functions linear in the perturbation. Lowered flat local components are denoted by
\begin{align}
  \theta_{mn} &\overset{\text{note}}{=}\eta_{m\mu}(\theta^\mu)_n, & c_m &\overset{\text{note}}{=}\eta_{m\mu}c^\mu.
\end{align}
The diffeomorphism gauge condition is
\begin{align}\label{eq:diff_gauge_function_section4}
  \mathcal{G}^\mu[\theta] &\overset{\text{def}}{=} \kappa\,\partial_\nu \left[ \eta^{\nu a}(\theta^\mu)_a + \eta^{\mu a}(\theta^\nu)_a \right], &  \mathcal{G}_\mu &\overset{\text{def}}{=}\eta_{\mu\nu}\mathcal{G}^\nu,
\end{align}
and the local Lorentz gauge condition is
\begin{align}\label{eq:lorentz_gauge_function_section4}
  \mathcal{Z}_{mn}[\theta] &\overset{\text{def}}{=} \kappa(\theta_{mn}-\theta_{nm}), & \mathcal{Z}_{mn} &=-\mathcal{Z}_{nm}.
\end{align}
The first condition fixes diffeomorphisms, while the second one fixes the antisymmetric local Lorentz part of the density vierbein perturbation. A derivative local Lorentz gauge could be chosen, but it would make the local Lorentz ghost kinetic operator differential. The algebraic gauge keeps the local Lorentz ghost sector simple and avoids higher-derivative ghost propagators. In the flat background we use
\begin{align}\label{eq:gauge_fixing_operators_section4}
  Y^{\mu\nu} &\overset{\text{def}}{=}\eta^{\mu\nu},&  Z^{mnrs} &\overset{\text{def}}{=} \frac12(\eta^{mr}\eta^{ns}-\eta^{ms}\eta^{nr}).
\end{align}

The BRST variation of the diffeomorphism gauge function is
\begin{align}\label{eq:brst_G_flat_section4}
  \begin{split}
    \delta_\text{B}\mathcal{G}^\mu = -\Box c^\mu + \kappa \, \partial_\nu \Bigg[& \eta^{\nu a} \left( c^\rho\partial_\rho(\theta^\mu)_a - (\theta^\rho)_a\partial_\rho c^\mu + \frac12(\theta^\mu)_a\partial_\rho c^\rho + \chi_a{}^b(\theta^\mu)_b \right) \\
      & + \eta^{\mu a} \left(  c^\rho\partial_\rho(\theta^\nu)_a - (\theta^\rho)_a\partial_\rho c^\nu + \frac12(\theta^\nu)_a\partial_\rho c^\rho + \chi_a{}^b(\theta^\nu)_b \right) \Bigg],
  \end{split}
\end{align}
where $\Box\overset{\text{def}}{=}\partial_\rho\partial^\rho$. The zeroth-order local Lorentz ghost contribution cancels because $\chi_{mn}$ is antisymmetric. Similarly,
\begin{align}\label{eq:brst_Z_flat_section4}
  \begin{split}
    \delta_\text{B}\mathcal{Z}_{mn} =& \partial_m c_n- \partial_n c_m- 2\chi_{mn} \\
    &+ \kappa \, \eta_{m\mu} \left( c^\rho\partial_\rho(\theta^\mu)_n - (\theta^\rho)_n\partial_\rho c^\mu + \frac12(\theta^\mu)_n\partial_\rho c^\rho + \chi_n{}^s(\theta^\mu)_s \right) \\
    &- \kappa \, \eta_{n\mu} \left( c^\rho\partial_\rho(\theta^\mu)_m - (\theta^\rho)_m\partial_\rho c^\mu + \frac12(\theta^\mu)_m\partial_\rho c^\rho +\chi_m{}^s(\theta^\mu)_s \right).
  \end{split}
\end{align}
The part of \eqref{eq:brst_Z_flat_section4} that does not contain $\theta$ produces a quadratic mixing between the diffeomorphism ghost and the local Lorentz antighost,
\begin{align}
  -\overline\chi^{mn}(\partial_m c_n-\partial_n c_m).
\end{align}
This mixing comes from the diffeomorphism variation of the antisymmetric part of the density vierbein perturbation and is not a pathology.

It is useful to diagonalize the ghost quadratic form by introducing the antisymmetric ghost
\begin{align}\label{eq:local_lorentz_ghost_redefinition_section4}
  c_{mn} &\overset{\text{def}}{=} \chi_{mn} - \frac12(\partial_m c_n-\partial_n c_m), & c_{mn} &= - c_{nm}.
\end{align}
The one-index ghost $c_m$ is the lowered diffeomorphism ghost, while the two-index ghost $c_{mn}$ is the redefined local Lorentz ghost. The antighost $\overline\chi^{mn}$ is not redefined. The change of variables is local, linear, and triangular in the ghost sector. It therefore introduces no field-dependent Jacobian in the functional integral.

The quadratic gauge-fixed action in the diagonal ghost variables is
\begin{align}\label{eq:quadratic_gauge_fixed_action_section4}
  \begin{split}
    S_2^\text{flat} \!=\!\!\! \int\!\!\dd^dx \Bigg[& - \frac12 \, \kappa^{d-2} B^{i_1}{}_{\alpha_1}{}^{a_1} (\mathscr{P}^{-1})_{i_1}{}^{\alpha_1}{}_{a_1}{}_{i_2}{}^{\alpha_2}{}_{a_2} B^{i_2}{}_{\alpha_2}{}^{a_2} + \frac12\kappa^{4-d} \partial_{i_1}(\theta^{\alpha_1})_{a_1} (\mathscr{P}_\text{GF})^{i_1}{}_{\alpha_1}{}^{a_1}{}^{i_2}{}_{\alpha_2}{}^{a_2} \partial_{i_2}(\theta^{\alpha_2})_{a_2} \\
      &+ \frac12\kappa^{4-d} (\theta^{\alpha_1})_{a_1} (\mathcal{P}_\text{L})_{\alpha_1}{}^{a_1}{}_{\alpha_2}{}^{a_2} (\theta^{\alpha_2})_{a_2} + \kappa^{4-d}\overline c_\mu\Box c^\mu + 2\kappa^{4-d}\overline\chi^{mn}c_{mn} \Bigg].
  \end{split}
\end{align}
Here
\begin{align}\label{eq:P_GF_definition_section4}
  \begin{split}
    \mathscr{P}_\text{GF} &\overset{\text{def}}{=} \mathscr{P}+\mathcal{P}_\text{diff}, \\
    (\mathcal{P}_\text{diff})^{i_1}{}_{\alpha_1}{}^{a_1}{}^{i_2}{}_{\alpha_2}{}^{a_2} &\overset{\text{def}}{=} \frac{1}{\epsilon} \Big( \eta_{\alpha_1\alpha_2}\eta^{i_1a_1}\eta^{i_2a_2} + \delta_{\alpha_1}^{a_2}\eta^{i_1a_1}\delta_{\alpha_2}^{i_2} + \delta_{\alpha_2}^{a_1}\delta_{\alpha_1}^{i_1}\eta^{i_2a_2} + \eta^{a_1a_2}\delta_{\alpha_1}^{i_1}\delta_{\alpha_2}^{i_2} \Big), \\
    (\mathcal{P}_\text{L})_{\alpha_1}{}^{a_1}{}_{\alpha_2}{}^{a_2} &\overset{\text{def}}{=} \frac{2}{\rho} \left( \eta_{\alpha_1\alpha_2}\eta^{a_1a_2} - \delta_{\alpha_1}^{a_2}\delta_{\alpha_2}^{a_1} \right).
  \end{split}
\end{align}

The interaction part of the reduced flat gauge-fixed action is the sum of the finite auxiliary field interactions from \eqref{eq:finite_perturbative_action_section3} and the ghost interactions obtained from \eqref{eq:brst_G_flat_section4} and \eqref{eq:brst_Z_flat_section4} after the change of variables \eqref{eq:local_lorentz_ghost_redefinition_section4}. In terms of $c_{mn}$, the latter can be written compactly as
\begin{align}\label{eq:ghost_interactions_section4}
  \begin{split}
    S_\text{gh,int}^\text{flat} \!\!=\! \kappa^{5-d}\!\!\!\int\!\!\dd^dx \Bigg\{\! (\partial_\nu\overline c_\mu) \Bigg[& \eta^{\nu a} \left( c^\rho\partial_\rho(\theta^\mu)_a - (\theta^\rho)_a\partial_\rho c^\mu + \frac12(\theta^\mu)_a\partial_\rho c^\rho + \left[c_a{}^b+\frac12(\partial_a c^b-\partial^b c_a)\right](\theta^\mu)_b \right) \\
      &\hspace{-4mm}+\eta^{\mu a} \left( c^\rho\partial_\rho(\theta^\nu)_a - (\theta^\rho)_a\partial_\rho c^\nu + \frac12(\theta^\nu)_a\partial_\rho c^\rho + \left[c_a{}^b+\frac12(\partial_a c^b-\partial^b c_a)\right](\theta^\nu)_b \right) \Bigg] \\
    & \hspace{-19mm} - \overline\chi^{mn} \Bigg[ \eta_{m\mu} \left( c^\rho\partial_\rho(\theta^\mu)_n - (\theta^\rho)_n\partial_\rho c^\mu + \frac12(\theta^\mu)_n\partial_\rho c^\rho + \left[c_n{}^s+\frac12(\partial_n c^s-\partial^s c_n)\right](\theta^\mu)_s \right) \\
      & \hspace{-14mm} - \eta_{n\mu} \left( c^\rho\partial_\rho(\theta^\mu)_m - (\theta^\rho)_m\partial_\rho c^\mu +\frac12(\theta^\mu)_m\partial_\rho c^\rho + \left[c_m{}^s+\frac12(\partial_m c^s-\partial^s c_m)\right](\theta^\mu)_s \right)\Bigg] \Bigg\}.
  \end{split}
\end{align}
Here $c_a{}^b\overset{\text{note}}{=}\eta^{bs}c_{as}$, $\partial^b\overset{\text{note}}{=}\eta^{b\rho}\partial_\rho$, and $c^b\overset{\text{note}}{=}\eta^{b\rho}c_\rho$. In particular, $c_m$ is the lowered diffeomorphism ghost, whereas $c_{mn}$ is the redefined local Lorentz ghost. The full reduced flat action is therefore
\begin{align}\label{eq:reduced_flat_action_section4}
  S_\text{red}^\text{flat} = S_2^\text{flat} + S_\text{H,int}^\text{flat} + S_\text{gh,int}^\text{flat},
\end{align}
where $S_\text{H,int}^\text{flat}$ is the interaction part of \eqref{eq:finite_perturbative_action_section3}.

The propagators are read from the quadratic action \eqref{eq:quadratic_gauge_fixed_action_section4}. For the explicit propagators displayed below, we choose the convenient gauge
\begin{align}
  \epsilon=\frac12, \qquad \rho=1.
\end{align}
In momentum space, with all momenta incoming and $\partial_\mu\to \ii \, p_\mu$, they are the inverses of the corresponding quadratic kernels. The complete $\theta$ propagator includes both a symmetric pole term and an antisymmetric algebraic term:
\begin{align}\label{eq:propagators_symbolic_section4}
  \begin{split}
    \left\langle 0 \middle| \mathrm{T}\!\left\{(\theta^\alpha)_a(p)(\theta^\beta)_b(-p)\right\}\middle|0\right\rangle &= \ii\,\kappa^{d-4}\Bigg[\frac{\eta^{\alpha\beta}\eta_{ab} + \delta^\alpha{}_b\delta^\beta{}_a - \delta^\alpha{}_a\delta^\beta{}_b}{8(p^2+\ii0)}+\frac18\left(\eta^{\alpha\beta}\eta_{ab}-\delta^\alpha{}_b\delta^\beta{}_a\right)\Bigg], \\
    \left\langle 0 \middle| \mathrm{T}\!\left\{B^{i_1}{}_{\alpha_1}{}^{a_1}(p)B^{i_2}{}_{\alpha_2}{}^{a_2}(-p)\right\}\middle|0\right\rangle &= -2i\,\kappa^{2-d}\Bigg[ -\eta^{i_1a_2}\delta_{\alpha_2}^{a_1}\delta_{\alpha_1}^{i_2}+\eta^{i_1i_2}\delta_{\alpha_1}^{a_2}\delta_{\alpha_2}^{a_1}-\eta^{i_1a_2}\eta^{i_2a_1}\eta_{\alpha_1\alpha_2}\\
      & \hspace{60pt} -\eta^{i_2a_1}\delta_{\alpha_1}^{a_2}\delta_{\alpha_2}^{i_1} +\eta^{a_1a_2}\eta^{i_1i_2}\eta_{\alpha_1\alpha_2} -2\eta^{a_1a_2}\delta_{\alpha_1}^{i_1}\delta_{\alpha_2}^{i_2} \\
      & \hspace{60pt} +\eta^{a_1a_2}\delta_{\alpha_2}^{i_1}\delta_{\alpha_1}^{i_2} -\frac{2}{d-2}\eta^{i_1i_2} \delta_{\alpha_1}^{a_1}\delta_{\alpha_2}^{a_2} \Bigg],\\
    \left\langle 0 \middle| \mathrm{T}\!\left\{c^\mu(p)\overline c_\nu(-p)\right\}\middle|0\right\rangle &= \ii\,\kappa^{d-4}\,\frac{1}{-p^2+\ii0}\, \delta^\mu_\nu,\\
    \left\langle 0 \middle| \mathrm{T}\!\left\{c_{mn}(p)\overline\chi_{rs}(-p)\right\}\middle|0\right\rangle &= \ii\,\kappa^{d-4}~ \cfrac14\,(\eta_{mr}\eta_{ns}-\eta_{ms}\eta_{nr}).
  \end{split}
\end{align}
For arbitrary nonzero $\rho$ at $\epsilon=1/2$, the last term of the $\theta$ propagator is multiplied by $\rho$. It contracts the antisymmetric local Lorentz gauge component and carries no physical pole. The sharp gauge is the limit $\rho\to0$ of the propagators, not a direct substitution into the reduced action containing $1/\rho$. The auxiliary propagator is defined on $\Img\OO_0$. It and the local Lorentz ghost propagator are algebraic.

It is useful to check if we can recover the well-known expression for the graviton propagator. The standard propagator describes the metric perturbation $g_{\mu\nu}=\eta_{\mu\nu}+\kappa h_{\mu\nu}$. With $\theta^{\mu\nu}=\eta^{\nu a}(\theta^\mu)_a$ and~\eqref{eq:density_vierbein_factorisation} we obtain the relation between $h_{\mu\nu}$ and density vierbein perturbations $\theta$ at the linear order:
\begin{align}\label{eq:theta_metric_linear_relation}
  h^{\mu\nu}=-2\theta^{(\mu\nu)}+\frac{2}{d-2}\eta^{\mu\nu}\theta^\rho{}_\rho.
\end{align}
Applying this relation to both legs of the propagator, we obtain
\begin{align}\label{eq:metric_propagator_check}
  \langle h^{\mu\nu}(p)h^{\rho\sigma}(-p)\rangle =\kappa^{d-4} \, \frac{\ii}{(p^2+\ii0)} ~\cfrac12\, \left[\eta^{\mu\rho}\eta^{\nu\sigma}+\eta^{\mu\sigma}\eta^{\nu\rho} -\frac{2}{d-2}\eta^{\mu\nu}\eta^{\rho\sigma}\right].
\end{align}

The model now has six interaction terms involving the following fields
\begin{align}
  \theta^3, \qquad \theta^4, \qquad \theta^2B, \qquad \theta^3B, \qquad \theta BB, \qquad \theta^2BB.
\end{align}
The ghost vertices are obtained from \eqref{eq:ghost_interactions_section4}. Their number is also finite because the flat-background BRST transformation \eqref{eq:brst_theta_flat_section4} is at most linear in $\theta$ while the gauge-fixing functions \eqref{eq:diff_gauge_function_section4} and \eqref{eq:lorentz_gauge_function_section4} are linear in $\theta$. Thus, the ghost sector does not reintroduce an infinite tower of interactions. The corresponding momentum-space coefficients are collected in Appendix~\ref{app:feynman-rules}.

Momentum-space coefficients follow by replacing derivatives with $\ii p_\mu$, taking all momenta incoming, and differentiating the specified polynomial representative. Equation~\eqref{eq:reduced_flat_action_section4} determines these coefficients while the nonlinear auxiliary restriction remains part of the classical formulation. The free $B$ kernel is algebraic, the Nakanishi--Lautrup fields have been eliminated, and the ghost fields belong to the gauge-fixing sector. None supplies additional physical graviton polarizations in the background quadratic analysis.

The gauge parameters $\epsilon$ and $\rho$ control the coordinate and local Lorentz gauges. Off-shell Green functions and individual Feynman rules depend on them, while physical on-shell quantities should not. The output of this section is the background quadratic structure and a finite set of interaction coefficients for the constrained gravitational polynomial and its gauge-fixing sector. Their use beyond the free theory retains the nonlinear auxiliary restriction. The next question is whether this finite structure survives when the theory is coupled to Dirac fermions, where the spin connection enters explicitly.

\section{Coupling to Dirac Fermions}\label{sec:Dirac_action}

The previous sections show that the pure Hilbert action admits a finite perturbative formulation in density vierbein variables. The next test is the coupling to matter fields that intrinsically require a local Lorentz frame. Dirac fermions provide the most direct example. Their coupling to gravity cannot be expressed naturally in terms of metric variables alone, because the spinor representation is a representation of the local Lorentz group. This is the reason why the vierbein formulation is not merely a convenient rewriting of the metric theory, but the natural setting for the fermion-gravity coupling.

We work locally with ordinary spinors of density weight zero and the torsion-free connection fixed in Appendix~\ref{app:vierbein-conventions}. The Dirac adjoint is $\overline\psi=\psi^\dagger\gamma^0$. We use constant local Lorentz gamma matrices satisfying
\begin{align}
  \{\gamma^a,\gamma^b\}=2\eta^{ab}.
\end{align}
The vierbein converts spacetime indices into local Lorentz indices and allows these gamma matrices to be used at each spacetime point. With the spinor Lorentz generators $\Sigma^{ab}=-\tfrac{\ii}{4}[\gamma^a,\gamma^b]$, the spin connection matrix acting on spinors is
\begin{align}
  \omega_\mu \overset{\text{def}}{=} -\frac{\ii}{2}(\omega_\mu)_{ab}\Sigma^{ab} = -\frac14(\omega_\mu)_{ab}\gamma^a\gamma^b .
\end{align}
In the conventions used here, the covariant derivatives of the spinor and of the conjugated spinor are
\begin{align}
  \nabla_\mu\psi &= \pd_\mu\psi-\omega_\mu\psi, &
  \nabla_\mu\overline\psi &= \pd_\mu\overline\psi+\overline\psi\,\omega_\mu .
\end{align}
The spin connection with two lower local Lorentz indices is antisymmetric,
\begin{align}
  (\omega_\mu)_{ab}=-(\omega_\mu)_{ba},
\end{align}
which follows from the spin connection preserving the local Lorentz metric. The sign of the matrix connection is fixed by compatibility with the vector connection used throughout the paper. In particular,
\begin{align}\label{eq:spinor_gamma_compatibility_section5}
 [\omega_\mu,\gamma^a]=(\omega_\mu)^a{}_b\gamma^b.
\end{align}
Thus $\overline\psi\gamma^a\psi$ has the covariant derivative of an upper Lorentz vector. The spinor connection uses the ordinary antisymmetric connection $\omega_{\mu ab}$, rather than the trace-shifted density connection $\Omega_{\mu ab}$.

The minimally coupled Dirac action can first be written in the non-symmetric form
\begin{align}\label{eq:Dirac_action_nonsymmetric_section5}
  S_\text{D} = \int\dd^dx\sqrt{-g}\left[ \ii\,\overline\psi\gamma^m(e^\mu)_m\nabla_\mu\psi-m\overline\psi\psi  \right].
\end{align}
Define $\gamma^\mu(x)=(e^\mu)_m\gamma^m$. The tetrad postulate and~\eqref{eq:spinor_gamma_compatibility_section5} imply
\begin{align}\label{eq:dirac_current_divergence_section5}
 \frac1e\partial_\mu\big(e\,\overline\psi\gamma^\mu\psi\big)
 =(\nabla_\mu\overline\psi)\gamma^\mu\psi+\overline\psi\gamma^\mu\nabla_\mu\psi.
\end{align}
Consequently, the non-symmetric action differs from the following symmetric action by the boundary term and reads:
\begin{align}\label{eq:Dirac_action_symmetric_section5}
  S_\text{D} = \int\dd^dx\sqrt{-g}\left[ \frac{\ii}{2} \left( \overline\psi\gamma^m(e^\mu)_m\nabla_\mu\psi - \nabla_\mu\overline\psi\,\gamma^m(e^\mu)_m\psi \right) - m\overline\psi\psi \right].
\end{align}
This is the form in which the spin connection contribution is most transparent.

The derivative part of \eqref{eq:Dirac_action_symmetric_section5} does not couple directly to the density vierbein $(E^\mu)_m$. With our conventions,
\begin{align}
  (E^\mu)_m &\overset{\text{def}}{=}\sqrt e\,(e^\mu)_m, & e &= \sqrt{-g}, & \sqrt{-g}\,(e^\mu)_m &=e\,(e^\mu)_m=\sqrt e\,(E^\mu)_m .
\end{align}
It is therefore useful to introduce a new rescaled vierbein
\begin{align}\label{eq:calE_definition_section5}
  (\mathcal E^\mu)_m \overset{\text{def}}{=} \sqrt{-g}\,(e^\mu)_m = \sqrt e\,(E^\mu)_m .
\end{align}
The derivative part of the Dirac action becomes
\begin{align}\label{eq:Dirac_derivative_calE_section5}
  \frac{\ii}{2} \int\dd^dx\,(\mathcal E^\mu)_m \left[ \overline\psi\gamma^m\pd_\mu\psi - \pd_\mu\overline\psi\,\gamma^m\psi  \right].
\end{align}
The mass term is
\begin{align}\label{eq:Dirac_mass_density_section5}
  S_{\rm mass}=-m\int\dd^dx\,(\det E)^{2/(d-2)}\overline\psi\psi.
\end{align}
In four dimensions, its density factor is $\det E$.

The transition from $E$ to $\mathcal{E}$ is an invertible local field redefinition since
\begin{align}
  \det E &= e^{\frac{d-2}{2}} \, , &  \det\mathcal{E} &= e^{d-1},
\end{align}
which results in
\begin{align}\label{eq:E_to_calE_map_section5}
  (\mathcal{E}^\mu)_m &= (\det E)^{\frac{1}{d-2}}(E^\mu)_m \, , & (E^\mu)_m &= (\det\mathcal{E})^{-\frac{1}{2(d-1)}}(\mathcal{E}^\mu)_m .
\end{align}
Although this field redefinition is locally invertible on the chosen nondegenerate branch, it need not preserve polynomial dependence on the perturbations. We therefore retain $E$ as the gravitational variable and use $\mathcal E$ only as shorthand for the coefficient in~\eqref{eq:Dirac_derivative_calE_section5}.

Substituting the spinor covariant derivatives into \eqref{eq:Dirac_action_symmetric_section5} separates the action into the derivative part, the mass term, and the spin connection contribution. If
\begin{align}
  \omega_m\overset{\text{note}}{=}(e^\mu)_m\omega_\mu,
\end{align}
then the spin connection contribution is
\begin{align}\label{eq:Dirac_spin_connection_matrix_section5}
  \mathcal{S}_\omega = -\frac{\ii}{2} \int\dd^dx\sqrt{-g}\, \overline\psi\left(\gamma^m\omega_m+\omega_m\gamma^m\right)\psi .
\end{align}
This term is the part of the Dirac action that is most sensitive to the choice of gravitational variables.

The expression \eqref{eq:Dirac_spin_connection_matrix_section5} follows directly from the two spin connection terms produced by the symmetric action. The first term in \eqref{eq:Dirac_action_symmetric_section5} contributes $\overline\psi\gamma^m\omega_m\psi$, while the second contributes $\overline\psi\omega_m\gamma^m\psi$, with the common coefficient $-\ii/2$ and the volume density. Thus the spin connection contribution is proportional to the anticommutator $\{\gamma^m,\omega_m\}$ and must be retained in the fermion-gravity Feynman rules.

Using the definition of $\omega_m$ and the antisymmetry of $(\omega_m)_{ab}$, one obtains
\begin{align}\label{eq:gamma_omega_anticommutator_section5}
  \{\gamma^m,\omega_m\} = -\frac12(\omega_m)_{ab}\gamma^{mab} =\frac12(\omega_m)_{ab}\gamma^a\gamma^m\gamma^b,
\end{align}
where $\gamma^{mab}\overset{\text{def}}{=}\gamma^{[m}\gamma^a\gamma^{b]}$ denotes antisymmetrization with weight $1/3!$. The identity follows from
\begin{align}
  \{\gamma^m,\gamma^a\gamma^b\}=2\gamma^{mab}+2\eta^{ab}\gamma^m,
\end{align}
whose second term vanishes after contraction with $(\omega_m)_{ab}$. Equivalently, antisymmetry in $a,b$ gives
\begin{align}
  (\omega_m)_{ab}\gamma^a\gamma^m\gamma^b=-(\omega_m)_{ab}\gamma^{mab}.
\end{align}
Here
\begin{align}
  (\omega_m)_{ab}\overset{\text{note}}{=}(e^\mu)_m(\omega_\mu)_{ab}.
\end{align}
Consequently, \eqref{eq:Dirac_spin_connection_matrix_section5} becomes
\begin{align}\label{eq:Dirac_spin_connection_term_section5}
  \mathcal{S}_\omega = +\frac{\ii}{4} \int\dd^dx\sqrt{-g}\, (\omega_m)_{ab}\, \overline\psi\gamma^{mab}\psi .
\end{align}

Expression \eqref{eq:Dirac_spin_connection_term_section5} shows that we can introduce a new variable
\begin{align}\label{eq:spin_connection_density_section5}
  \Xi_{mab} \overset{\text{def}}{=} \sqrt{-g}\,(\omega_m)_{ab}.
\end{align}
In terms of this object, the spin connection part of the Dirac action is linear
\begin{align}\label{eq:Dirac_spin_connection_density_term_section5}
  \mathcal{S}_\omega = +\frac{\ii}{4} \int\dd^dx\, \Xi_{mab}\, \overline\psi\gamma^{mab}\psi .
\end{align}
Thus the spin connection term couples to the totally antisymmetric component $\Xi_{[mab]}$. Together with~\eqref{eq:Dirac_derivative_calE_section5} and~\eqref{eq:Dirac_mass_density_section5}, this gives the three contributions to the standard Dirac action in density vierbein variables. For completeness, we can also write the connection term entirely in terms of the density vierbein. The anholonomy convention of Appendix~\ref{app:vierbein-conventions} gives
\begin{align}\label{eq:dirac_anholonomy_contraction_section5}
  (\omega_m)_{ab}\gamma^{mab} =-\frac12 C_{abm}\gamma^{mab} =(e^\alpha)_a(e^\beta)_b\partial_\alpha(e_\beta)_m\gamma^{mab}.
\end{align}
Substitute $(e^\alpha)_a=e^{-1/2}(E^\alpha)_a$ and $(e_\beta)_m=e^{1/2}(E_\beta)_m$. The derivative of $\sqrt e$ produces a term proportional to $\eta_{bm}\gamma^{mab}=0$. Hence
\begin{align}\label{eq:dirac_density_connection_section5}
  \Xi_{mab}\gamma^{mab} =(\det E)^{1/(d-2)}(E^\alpha)_a(E^\beta)_b\partial_\alpha(E_\beta)_m\gamma^{mab}.
\end{align}
Combining the three contributions yields the exact action
\begin{align}\label{eq:dirac_action_density_complete_section5}
  \begin{split}
    S_{\rm D}=\int\dd^dx\,\Bigg\{&\frac{\ii}{2}(\det E)^{1/(d-2)}(E^\mu)_m \big[\overline\psi\gamma^m\partial_\mu\psi-(\partial_\mu\overline\psi)\gamma^m\psi\big]\\
    &+\frac{\ii}{4}(\det E)^{1/(d-2)}(E^\alpha)_a(E^\beta)_b \partial_\alpha(E_\beta)_m\,\overline\psi\gamma^{mab}\psi-m(\det E)^{2/(d-2)}\overline\psi\psi\Bigg\}.
  \end{split}
\end{align}

In $d=4$, the derivative coefficient is $\sqrt{\det E}(E^\mu)_m$ and it is an infinite series in $\kappa$. The connection and mass terms contain no derivatives of the spinors and cannot cancel this coefficient. Thus both massive and massless standard Dirac fields have an infinite tower of density vierbein interactions in this representation. The four-dimensional mass term itself is polynomial, of degree at most four in $\theta$.

\section{Other Matter Couplings}\label{sec:matter_couplings}

We now examine which additional matter couplings retain a finite expansion in the density vierbein perturbation. Throughout this section, we consider only the $d=4$ case since the original Horndeski theory is defined in that dimension. Let us closely examine
\begin{align}\label{eq:matter_density_identities}
  E & = \det(E^\mu)_a, & \mathfrak g^{\mu\nu}&=\eta^{ab}(E^\mu)_a(E^\nu)_b, & g^{\mu\nu}&=\frac{\mathfrak g^{\mu\nu}}{E}.
\end{align}
The determinant $E=\det(\mathbf1+\kappa\theta)$ is a polynomial of degree four and $\mathfrak g^{\mu\nu}$ is also a polynomial of degree two. In contrast, $g^{\mu\nu}$ is non-polynomial since $E^{-1}$ is an infinite Taylor series in $\kappa$. These identities immediately highlight the scalar field potential and the canonical scalar kinetic term. In contrast, they suggest that the vector-field canonical kinetic term may not be polynomial, since it requires two inverse metrics. Below, we examine these cases in detail.

We begin with the scalar terms, for which we can count the determinant factors directly. Define the scalar kinetic invariant by
\begin{align}\label{eq:scalar_X_definition}
  X\overset{\text{def}}{=}\frac12g^{\mu\nu}\partial_\mu\phi\partial_\nu\phi.
\end{align}
For arbitrary functions $K(\phi)$ and $V(\phi)$,
\begin{align}\label{eq:finite_scalar_terms}
  \sqrt{-g}\,[K(\phi)X+V(\phi)]  =\frac12 K(\phi)\mathfrak g^{\mu\nu}\partial_\mu\phi\partial_\nu\phi+\Delta V(\phi).
\end{align}
The kinetic and potential terms have degrees two and four in $E$, respectively, so such a model has a finite number of interaction terms\footnote{We do not address whether $V(\phi)$ or $K(\phi)$ introduce an infinite number of scalar field self-interaction terms, for this is a completely different discussion.} with density vierbein perturbation.

To include derivative and non-minimal scalar couplings, we study the Horndeski gravity~\cite{Horndeski:1974wa,Kobayashi:2011nu}, which is given by the following action\footnote{It is important to note that the action in the original paper uses a different metric signature.}
\begin{align}\label{eq:Horndeski_action}
  \begin{split}
    &S_{\rm H}=\int d^4x\,\sqrt{-g}\Bigg[ G_2(\phi,X) + G_3(\phi,X)\Box\phi + G_4(\phi,X)R -G_{4,X}(\phi,X) \left[ (\Box\phi)^2 -\nabla_\mu\nabla_\nu\phi\,\nabla^\mu\nabla^\nu\phi \right] \\
      & + G_5(\phi,X)G_{\mu\nu}\nabla^\mu\nabla^\nu\phi +\cfrac{G_{5,X}(\phi,X)}{3!} \left[ (\Box\phi)^3 -3\Box\phi\, \nabla_\mu\nabla_\nu\phi\,\nabla^\mu\nabla^\nu\phi +2\nabla_\mu\nabla^\nu\phi\, \nabla_\nu\nabla^\rho\phi\, \nabla_\rho\nabla^\mu\phi \right] \Bigg].
  \end{split}
\end{align}
The functions $G_i=G_i(\phi,X)$ are assumed analytic in $X$ near zero, with sufficiently differentiable coefficient functions of $\phi$. The curvature coefficient includes the Hilbert term, and the general relativity corresponds to $G_4=-2/\kappa^2$ and $G_2=G_3=G_5=0$.

Before counting density vierbein powers in this larger class, we must fix the freedom to add surface terms. Otherwise, a non-polynomial term in one representative could be canceled by another term after integration by parts. Two analytic tuples $G_i$ and $g_i$ describe the same local bulk action precisely when their difference has the form
\begin{align}\label{eq:Horndeski_boundary_equivalence}
  \begin{split}
    G_2-g_2&=2\,A'(\phi)X+2f'''(\phi)X^2, \\
    G_3-g_3&=A(\phi)+3f''(\phi)X,\\
    G_4-g_4&=-f'(\phi)X, \\
    G_5-g_5&=f(\phi).
  \end{split}
\end{align}
Primes denote derivatives with respect to $\phi$. The term $A\Box\phi$ equals $-2a'X$ after integration by parts. For the remaining terms, define
\begin{align}
  J^\mu=\nabla^\mu\phi\,\Box\phi-\nabla_\nu\phi\,\nabla^\mu\nabla^\nu\phi.
\end{align}
The curvature convention in Appendix~\ref{app:vierbein-conventions} gives
\begin{align}\label{eq:Horndeski_current_identity}
  \nabla_\mu J^\mu = (\Box\phi)^2 - (\nabla_\mu\nabla_\nu\phi)(\nabla^\mu\nabla^\nu\phi) - R_{\mu\nu}\nabla^\mu\phi\,\nabla^\nu\phi.
\end{align}
Using this identity and the contracted Bianchi identity yields, modulo a surface term,
\begin{align}\label{eq:Horndeski_f_identity}
  \int\!\!\dd^4x\sqrt{-g}\,fG_{\mu\nu}\nabla^\mu\nabla^\nu\phi \!=\! \int\!\!\dd^4x\sqrt{-g}\,\Big\{ f'XR - f'\big[(\Box\phi)^2 - (\nabla_\mu\nabla_\nu\phi)(\nabla^\mu\nabla^\nu\phi)\big] -3f''X\Box\phi-2f'''X^2 \Big\}.
\end{align}
Equations~\eqref{eq:Horndeski_f_identity} and the identity for $A\Box\phi$ prove that~\eqref{eq:Horndeski_boundary_equivalence} leaves the action unchanged modulo a surface term.

We can now specify a representative for which we will examine the expansion. We first impose $G_5(\phi,0)=0$ and then $G_3(\phi,0)=0$ using~\eqref{eq:Horndeski_boundary_equivalence}, including the induced changes in $G_2$ and $G_4$.
With this prescription fixed, we can now identify which Horndeski models admit a finite number of interactions with the density vierbein variables. In this representation, the density vierbein expansion terminates if and only if the analytic Horndeski equivalence class contains a representative
\begin{align}\label{eq:finite_Horndeski_class}
  G_2 &= K(\phi)X+V(\phi), & G_3 &= 0, & G_4 &=-\frac2{\kappa^2}+F(\phi), & G_5 &= 0.
\end{align}

To see this, let us rescale $E\mapsto tE$ with constant $t>0$, holding matter fields fixed. Every derivative of $E$ also acquires one factor of $t$, so a finite polynomial in $E$ and its derivatives has only non-negative integer scaling degrees. In four dimensions,
\begin{align}\label{eq:matter_scaling_laws}
  \sqrt{-g} &\mapsto t^4\sqrt{-g}, & g^{\mu\nu} &\mapsto t^{-2}g^{\mu\nu}, & X &\mapsto t^{-2}X, & \Box\phi &\mapsto t^{-2}\Box\phi, & R &\mapsto t^{-2}R.
\end{align}
The Levi-Civita connection and $G_{\mu\nu}$ are unchanged. After normalizing the surface terms and separating the finite sectors $G_2(\phi,0)$, $G_{2,X}(\phi,0)X$, and $H(\phi)R$, the remaining Taylor monomials have the degrees in Table~\ref{tab:Horndeski_degrees}.
\begin{table}[htbp]
  \centering
  \begin{tabular}{lll}
    \hline
    Sector & Remaining monomials & Degree including $\sqrt{-g}$\\
    \hline
    $G_2$ & $X^n$, $n\ge2$ & $4-2n$\\
    $G_3$ & $X^n$, $n\ge1$ & $2-2n$\\
    $G_4$ & $X^n$, $n\ge1$ & $2-2n$\\
    $G_5$ & $X^n$, $n\ge1$ & $-2n$\\
    \hline
  \end{tabular}
  \caption{Scaling degrees of the residual analytic Horndeski sectors under $E\mapsto tE$. The $G_4$ and $G_5$ rows include their complete curvature and scalar-Hessian combinations.}
  \label{tab:Horndeski_degrees}
\end{table}

Group terms of equal degree before using the scaling argument. The density vierbein Euler--Lagrange derivative of an action of degree $w$ has degree $w-1$. 
Since all residual degrees satisfy $w\le0$, a polynomial representative requires the Euler--Lagrange derivative of each group to vanish: a polynomial derivative cannot have negative degree. This comparison also holds for the convergent analytic series. Setting $z=t^{-1}$, the residual derivative has strictly positive powers of $z$ near zero, whereas a polynomial in the scaled density vierbein and its derivatives has only non-positive powers. Small scalar gradients give a common domain of convergence, and uniqueness of the power series forces each residual coefficient to vanish.

Invertibility of the vierbein then implies zero metric variation of each grouped action. The result proved above requires each such group to have the surface-term form~\eqref{eq:Horndeski_boundary_equivalence}. Its normalized $G_5(\phi,0)$ and $G_3(\phi,0)$ vanish, forcing $f=A=0$. Thus every residual group vanishes, including possible cancellations between distinct Horndeski sectors—this proves~\eqref{eq:finite_Horndeski_class} within the stated representation.

The preceding result settles the analytic Horndeski class within the stated prescription. It does not directly cover scalar coupling to the Gauss--Bonnet invariant, whose Horndeski representation is non-analytic. We therefore examine this coupling separately, using the same scaling criterion. In four dimensions define
\begin{align}\label{eq:scalar_GB_action}
  \mathcal G_{\rm GB}&=R_{\mu\nu\rho\sigma}R^{\mu\nu\rho\sigma}-4R_{\mu\nu}R^{\mu\nu}+R^2, & S_{\rm EsGB}&=\int\dd^4x\sqrt{-g}\,\chi(\phi)\mathcal G_{\rm GB}.
\end{align}
For constant $\chi$, the term is locally a surface term, although its integral can retain topological and boundary information. A non-constant coupling has a nonzero metric variation in general.

The Horndeski representation of this coupling contains logarithms~\cite{Kobayashi:2011nu}. In the convention~\eqref{eq:Horndeski_action}, one representative is
\begin{align}\label{eq:scalar_GB_Horndeski_functions}
  \begin{split}
    G_2^{\rm EsGB}&=8\chi^{(4)}X^2 \left( 3- \ln\left|\frac{X}{X_0}\right| \right), \\
    G_3^{\rm EsGB}&=4\chi^{(3)}X \left( 7-3\ln\left|\frac{X}{X_0}\right| \right),\\
    G_4^{\rm EsGB}&=-4\chi''X \left( 2-\ln\left|\frac{X}{X_0}\right| \right), \\
    G_5^{\rm EsGB}&=-4\chi' , 
  \end{split}
\end{align}
where $X_0\neq0$ has the same dimension as $X$. These functions describe the Gauss--Bonnet sector and do not include the Hilbert term. Their non-analyticity at $X=0$ prevents direct use of~\eqref{eq:finite_Horndeski_class}.

The scaling argument applies directly to~\eqref{eq:scalar_GB_action}. Under~\eqref{eq:matter_scaling_laws}, $\mathcal G_{\rm GB}\mapsto t^{-4}\mathcal G_{\rm GB}$, so its density has degree zero. For $\chi'\neq0$, the density vierbein Euler--Lagrange derivative is nonzero in general and has degree $-1$. It therefore cannot arise from a finite polynomial density, even after integrations by parts. The scalar--Gauss--Bonnet coupling has an infinite density vierbein expansion in the metric representation used here, with no additional auxiliary fields for this sector.

We next turn from scalar--tensor interactions to vector fields. Here the determinant identities~\eqref{eq:matter_density_identities} expose the non-polynomial dependence directly, without requiring the surface-term analysis used for Horndeski theory. For a spacetime vector $A_\mu$ with $F_{\mu\nu}=\partial_\mu A_\nu-\partial_\nu A_\mu$, the Maxwell action becomes
\begin{align}\label{eq:Maxwell_density_action}
  S_{\rm M}=-\frac14\int\dd^4x\, \frac{\mathfrak g^{\mu\rho}\mathfrak g^{\nu\sigma}}{E} F_{\mu\nu}F_{\rho\sigma}.
\end{align}
The denominator is the density vierbein variable determinant. The determinant itself is a finite polynomial, but since it is present in the denominator, the whole expression is a Taylor series.

It is worth noting that the Proca mass term has a different dependence:
\begin{align}\label{eq:Proca_mass_density_action}
  S_{\rm Proca,mass}=\frac{M^2}{2}\int\dd^4x\,  \mathfrak g^{\mu\nu}A_\mu A_\nu.
\end{align}
It is quadratic in density vierbein variables, but the full Proca theory retains the Maxwell kinetic term, so it still admits an infinite tower of interaction terms.

For $SU(N)$ Yang--Mills theory one replaces the field strength in~\eqref{eq:Maxwell_density_action} by
\begin{align}
  F^a_{\mu\nu}=\partial_\mu A^a_\nu-\partial_\nu A^a_\mu + {\rm g} f^{abc}A^b_\mu A^c_\nu
\end{align}
and sum over the adjoint index. The same metric coefficient occurs, so the theory admits an infinite tower of interactions as well.

Combining these vector results with the scalar terms above and the fermion analysis of Section~\ref{sec:Dirac_action} determines the behavior of the minimally coupled Standard Model in its conventional matter variables. Its gauge kinetic terms retain the inverse determinant in~\eqref{eq:Maxwell_density_action}. The ordinary chiral fermion kinetic terms also carry the factor~$\sqrt E$ identified for Dirac fields in Section~\ref{sec:Dirac_action}. By contrast, the Higgs kinetic term is proportional to $\mathfrak g^{\mu\nu}(D_\mu H)^\dagger D_\nu H$, while the Higgs potential and Yukawa terms carry the polynomial density $E$. These finite pieces do not cancel the gauge or fermion kinetic coefficients. Thus the full Standard Model has an infinite tower of density vierbein interactions in the representation considered here.

Table~\ref{tab:matter_classification} summarizes the results. It lists the studied sectors rather than an unrestricted classification of all matter theories. The distinction between a finite density vierbein expansion and a finite number of all interaction vertices is essential when coefficient functions have arbitrary matter-field dependence.
\begin{table}[htbp]
  \centering
  \small
  \renewcommand{\arraystretch}{1.15}
  \begin{tabular}{p{0.40\textwidth}p{0.15\textwidth}p{0.35\textwidth}}
    \hline
    Matter sector & Finite in $\theta$? & Reason or qualification\\
    \hline
    $\sqrt{-g}\,V(\phi)$ & Yes & Degree at most four\\
    $\sqrt{-g}\,K(\phi)X$ & Yes & Degree at most two\\
    $\sqrt{-g}\,F(\phi)R$, & Yes & Degree at most two\\
    Analytic Horndeski theory & Yes & Degree at most four \\
    Scalar--Gauss--Bonnet, $\chi'\neq0$ & No & Nontrivial degree-zero action\\
    Maxwell and Yang--Mills kinetic terms & No & Factor $E^{-1}$\\
    Proca mass term alone & Yes & Degree at most two\\
    Complete Proca theory & No & Maxwell kinetic term\\
    Standard Dirac theory, with or without mass & No & $\sqrt E$ in its derivative coefficient\\
    Minimally coupled Standard Model & No & Gauge and fermion kinetic sectors\\
    \hline
  \end{tabular}
  \caption{Termination of the four-dimensional classical density vierbein expansion at fixed matter fields. The non-minimal curvature and analytic Horndeski results retain the prescribed gravitational auxiliary field on a patch where the total curvature coefficient is nonzero. We use no additional matter auxiliary fields.}
  \label{tab:matter_classification}
\end{table}

\section{Discussion}\label{sec:conclusions}

The density vierbein formulation extends the inverse metric density description of polynomial gravity~\cite{Cheung:2017kzx} to an explicit local Lorentz frame. The central result is the classical auxiliary field BRST formulation \eqref{eq:finite_perturbative_action_section3}, \eqref{eq:quadratic_gauge_fixed_action_section4}, \eqref{eq:ghost_interactions_section4}, which is polynomial in density vierbein variables. The kernel of the Hilbert quadratic form determines the allowed auxiliary field configurations. The corresponding perturbation theory produces cubic and quartic interaction classes. The construction thus yields a finite polynomial representation in density vierbein variables, alongside the cubic inverse-metric-density formulation.

The frame description also makes the coordinate and local Lorentz gauge sectors explicit. The BRST variations of the density vierbein and gauge-fixing fields determine both ghost sectors. A triangular ghost redefinition separates their background quadratic terms, and the chosen linear gauge functions give four cubic ghost interaction classes with six derivative structures. The density vierbein propagator includes a symmetric pole contribution and an antisymmetric algebraic contact term. These results specify the background quadratic structure and the momentum-space coefficients of the constrained action.

The matter analysis identifies how much finite dependence on the gravitational perturbation survives coupling to other fields. In four dimensions, scalar potentials and canonical scalar kinetic terms are polynomial in the density vierbein. Within the analytic Horndeski class and the prescribed auxiliary treatment, the finite family is represented by $G_2=K(\phi)X+V(\phi)$, $G_3=G_5=0$, and $G_4=-2/\kappa + F(\phi)$, modulo a corresponding surface term. Arbitrary scalar-dependent coefficients do not alter the bound on density vierbein valence.

The other matter models do not admit a finite number of couplings to density vierbein variables. Einstein--scalar--Gauss--Bonnet model, vector field kinetic term, and ordinary fermion kinetic terms produce infinite towers of interactions. The Proca mass term is polynomial, but its kinetic term is not. The minimally coupled Standard Model therefore remains plagued with the infinite tower through its gauge and fermion sectors, even though its Higgs kinetic term, potential, and Yukawa terms are polynomial in the density vierbein.

The scope of these statements is classical and representation-dependent. Polynomiality of the constrained density does not by itself provide a complete diagrammatic prescription in independent fields. Any such prescription must implement the nonlinear auxiliary restriction. We reserve its implementation, the functional measure, and applications for subsequent work. The present analysis does not classify unrestricted matter-field redefinitions or additional auxiliary constructions, and makes no claim of improved ultraviolet behavior or perturbative renormalizability. It establishes the constrained polynomial formulation, its background and gauge-fixing structure, and the matter couplings for which finite density vierbein valence is preserved.

\appendix

\section{Vierbein and Curvature Conventions}
\label{app:vierbein-conventions}

This appendix fixes the conventions for the ordinary vierbein, spin connection, torsion, and curvature used in the paper. The density vierbein variables and the density spin connection are treated separately in Appendix~\ref{app:density-spin connection}. Our conventions follow the standard vierbein formulation of general relativity~\cite{Misner:1973prb,Wald:1984rg,Carroll:2004stg,Nakahara:2003nw,Ortin:2004ms}.

We use the mostly minus local Lorentz metric
\begin{align}
  \eta_{ab}=\operatorname{diag}(+1,-1,\ldots,-1).
\end{align}
Greek letters $\mu,\nu,\rho,\ldots$ denote spacetime indices, while Latin letters $a,b,c,\ldots$ denote local Lorentz indices. Spacetime indices are raised and lowered with $g_{\mu\nu}$ and $g^{\mu\nu}$, and local Lorentz indices are raised and lowered with $\eta_{ab}$ and $\eta^{ab}$. We use
\begin{align}
  g &\overset{\text{def}}{=} \det g_{\mu\nu}, & e &\overset{\text{def}}{=} \det(e_\mu)^a, & e&= +\sqrt{-g},
\end{align}
where the last equality fixes the orientation. We also use the notation
\begin{align}
  \pd_\mu &\overset{\text{note}}{=}\frac{\partial}{\partial x^\mu}, & (\Gamma_\mu)^\rho{}_\nu & \overset{\text{note}}{=}\Gamma^\rho{}_{\mu\nu}.
\end{align}

The vierbein and inverse vierbein are denoted by $(e_\mu)^a$ and $(e^\mu)_a$ respectively. They satisfy
\begin{align}
  (e_\mu)^a(e^\mu)_b &= \delta^a_b, & (e_\mu)^a(e^\nu)_a &= \delta^\nu_\mu.
\end{align}
The spacetime metric is reconstructed from the vierbein by
\begin{align}
  g_{\mu\nu}&=(e_\mu)^a(e_\nu)^b\eta_{ab}, & g^{\mu\nu}&=(e^\mu)_a(e^\nu)_b\eta^{ab}.
\end{align}
These formulae also imply the determinant relation $e=+\sqrt{-g}$ with the orientation chosen above.

We use the numerical Levi-Civita symbols fixed by
\begin{align}
  \epsilon_{0\,1\,\cdots\,d-1} &\overset{\text{def}}{=} +1, & \epsilon^{0\,1\,\cdots\,d-1} &=(-1)^{d-1}.
\end{align}
The second relation follows from raising the local Lorentz indices with the mostly-minus metric $\eta^{ab}$.

Under a passive coordinate transformation
\begin{align}
  x^\mu &\mapsto x'^\mu(x), & \mathcal{J}_\mu{}^\nu &\overset{\text{def}}{=}\frac{\partial x^\nu}{\partial x'^\mu},
\end{align}
the vierbein transforms as a spacetime covector,
\begin{align}
  (e_\mu)^a\to \mathcal{J}_\mu{}^\nu(e_\nu)^a.
\end{align}
Under a local Lorentz transformation, it transforms as
\begin{align}
  (e_\mu)^a &\to (e_\mu)^b(J^{-1})_b{}^a, & J^a{}_cJ^b{}_d\eta_{ab}&=\eta_{cd}.
\end{align}
For an infinitesimal local Lorentz transformation, we use
\begin{align}
  \lambda_{ab} &=-\lambda_{ba}, & \delta V_a &= +\lambda_a{}^bV_b, & \delta V^a &= -V^b\lambda_b{}^a.
\end{align}
These signs fix the local Lorentz convention used in Section~\ref{sec:Hilbert_action}.

The covariant derivative acts on spacetime and local Lorentz indices simultaneously. Our sign conventions are
\begin{align}
  \begin{split}
    \nabla_\mu V_\nu{}^a &= \pd_\mu V_\nu{}^a - (\Gamma_\mu)^\rho{}_\nu V_\rho{}^a + (\omega_\mu)^a{}_b V_\nu{}^b, \\
    \nabla_\mu W^\nu{}_a &= \pd_\mu W^\nu{}_a + (\Gamma_\mu)^\nu{}_\rho W^\rho{}_a - W^\nu{}_b(\omega_\mu)^b{}_a.
  \end{split}
\end{align}
Here $(\Gamma_\mu)^\rho{}_\nu$ is the affine connection and $(\omega_\mu)^a{}_b$ is the ordinary spin connection.

The tetrad postulate is the condition
\begin{align}
  \nabla_\mu(e_\nu)^a=0.
\end{align}
In components, with the covariant derivative convention above, it reads
\begin{align}\label{eq:appendix_tetrad_postulate_expanded}
  \pd_\mu(e_\nu)^a - (\Gamma_\mu)^\rho{}_\nu(e_\rho)^a + (\omega_\mu)^a{}_b(e_\nu)^b =0.
\end{align}
It expresses compatibility between spacetime parallel transport and local Lorentz parallel transport.

The tetrad postulate, together with local Lorentz metric compatibility, implies spacetime metric compatibility. Indeed, if
\begin{align}
  \nabla_\mu\eta_{ab} &= 0, & \nabla_\rho(e_\mu)^a &= 0,
\end{align}
then
\begin{align}
  \nabla_\rho g_{\mu\nu} = \nabla_\rho\big[(e_\mu)^a(e_\nu)^b\eta_{ab}\big] =0.
\end{align}
Thus, metric compatibility may be expressed either in spacetime form or in local Lorentz form together with the tetrad postulate.

Local Lorentz metric compatibility also fixes the symmetry of the ordinary spin connection. Since
\begin{align}
  \nabla_\mu\eta_{ab} = - (\omega_\mu)^c{}_a\eta_{cb} - (\omega_\mu)^c{}_b\eta_{ac} =0,
\end{align}
one obtains
\begin{align}
  (\omega_\mu)_{ab} &= -(\omega_\mu)_{ba}, & (\omega_\mu)_{ab} &\overset{\text{def}}{=} \eta_{ac}(\omega_\mu)^c{}_b.
\end{align}
This antisymmetry is the property that distinguishes the ordinary spin connection from the density spin connection introduced in Section~\ref{sec:Hilbert_action}.

Solving the tetrad postulate~\eqref{eq:appendix_tetrad_postulate_expanded} for the ordinary spin connection gives
\begin{align}\label{eq:appendix_spin_connection_from_tetrad_postulate}
  (\omega_\mu)^a{}_b = (\Gamma_\mu)^\rho{}_\nu(e_\rho)^a(e^\nu)_b - \pd_\mu(e_\nu)^a(e^\nu)_b.
\end{align}
In the torsion-free second-order formulation used in this paper, the spin connection is therefore not an independent field; the vierbein determines it through the Levi-Civita connection.

Our coordinate torsion convention is
\begin{align}
  T^\rho{}_{\mu\nu} \overset{\text{def}}{=} (\Gamma_\mu)^\rho{}_{\nu}  - (\Gamma_\nu)^\rho{}_{\mu}.
\end{align}
The corresponding local Lorentz torsion two-form is defined by
\begin{align}
  e^a &\overset{\text{def}}{=}(e_\mu)^a\,\dd x^\mu, & \omega^a{}_b &\overset{\text{def}}{=}(\omega_\mu)^a{}_b\,\dd x^\mu, &  T^a &\overset{\text{def}}{=}\dd e^a+\omega^a{}_b\wedge e^b.
\end{align}
In components, this gives
\begin{align}
  T^a{}_{\mu\nu} = \pd_\mu(e_\nu)^a - \pd_\nu(e_\mu)^a + (\omega_\mu)^a{}_b(e_\nu)^b - (\omega_\nu)^a{}_b(e_\mu)^b.
\end{align}
Using the tetrad postulate, one obtains
\begin{align}
  T^a{}_{\mu\nu}=(e_\rho)^aT^\rho{}_{\mu\nu}.
\end{align}
Therefore, $T^\rho{}_{\mu\nu}=0$ is equivalent to $T^a=0$ once the tetrad postulate relates the affine connection and the spin connection.

In pseudo-Riemannian geometry, the Levi-Civita connection is the unique connection satisfying
\begin{align}
  \nabla_\rho g_{\mu\nu} &= 0, & T^\rho{}_{\mu\nu} &= 0.
\end{align}
Its coefficients are
\begin{align}
  (\Gamma_\mu)^\rho{}_{\nu} = \cfrac12 \, g^{\rho\sigma} \left( \pd_\mu g_{\nu\sigma} + \pd_\nu g_{\mu\sigma} - \pd_\sigma g_{\mu\nu} \right).
\end{align}
This is the affine connection used throughout the paper unless stated otherwise.

The anholonomy coefficients of the ordinary vierbein are defined by
\begin{align}
  C_{ab}{}^c &\overset{\text{def}}{=} \left[(e^\mu)_a\pd_\mu(e^\nu)_b-(e^\mu)_b\pd_\mu(e^\nu)_a\right](e_\nu)^c, & C_{abc} &\overset{\text{def}}{=}\eta_{cd}C_{ab}{}^d.
\end{align}
Thus $[e_a,e_b]=C_{ab}{}^c e_c$, where $e_a=(e^\mu)_a\partial_\mu$. These coefficients equal the negative of $(e^\mu)_a(e^\nu)_b[\partial_\mu(e_\nu)^c-\partial_\nu(e_\mu)^c]$. They are antisymmetric with respect to the first two indices
\begin{align}
  C_{ab}{}^c=-C_{ba}{}^c.
\end{align}
If
\begin{align}
  (\omega_a)_{bc} \overset{\text{def}}{=} (e^\mu)_a(\omega_\mu)_{bc},
\end{align}
then the torsion-free spin connection is
\begin{align}\label{eq:appendix_spin_connection_anholonomy}
  (\omega_a)_{bc} = \cfrac12  \left( C_{bca} + C_{acb} - C_{abc} \right).
\end{align}
This expression follows from $T^a=0$ and the antisymmetry of $(\omega_a)_{bc}$ in the last two indices.

The commutator defines the curvature of the affine connection
\begin{align}
  [\nabla_\mu,\nabla_\nu]V^\rho = R_{\mu\nu}{}^\rho{}_\sigma V^\sigma
\end{align}
in the torsion-free case. Equivalently,
\begin{align}
  R_{\mu\nu}{}^\rho{}_\sigma = \pd_\mu\Gamma^\rho{}_{\nu\sigma} - \pd_\nu\Gamma^\rho{}_{\mu\sigma} + \Gamma^\rho{}_{\mu\lambda}\Gamma^\lambda{}_{\nu\sigma} - \Gamma^\rho{}_{\nu\lambda}\Gamma^\lambda{}_{\mu\sigma}.
\end{align}
The curvature of the ordinary spin connection is defined by
\begin{align}
  R_{\mu\nu}{}^a{}_b \overset{\text{def}}{=} \pd_\mu(\omega_\nu)^a{}_b - \pd_\nu(\omega_\mu)^a{}_b + (\omega_\mu)^a{}_c(\omega_\nu)^c{}_b - (\omega_\nu)^a{}_c(\omega_\mu)^c{}_b.
\end{align}
In differential-form notation,
\begin{align}
  R^a{}_b &\overset{\text{def}}{=} \dd\omega^a{}_b+\omega^a{}_c\wedge\omega^c{}_b, & R^a{}_b &= \cfrac12 \, R_{\mu\nu}{}^a{}_b\,\dd x^\mu\wedge\dd x^\nu.
\end{align}

The coordinate curvature and the local Lorentz curvature are related by
\begin{align}
  R_{\mu\nu}{}^a{}_b  = (e_\rho)^a(e^\sigma)_bR_{\mu\nu}{}^\rho{}_\sigma.
\end{align}
This relation follows from the commutation of covariant derivatives acting on the tetrad postulate.

The Ricci tensor and scalar curvature are defined by
\begin{align}
  R_{\mu\nu} & \overset{\text{def}}{=} R_{\rho\mu}{}^\rho{}_\nu, &  R &\overset{\text{def}}{=}g^{\mu\nu}R_{\mu\nu}.
\end{align}
Equivalently,
\begin{align}
  R=(e^\mu)_a(e^\nu)_bR_{\mu\nu}{}^{ab}.
\end{align}
With these conventions, the Hilbert action used in the paper is
\begin{align}
  S_\text{H} = - \cfrac{2}{\kappa^{d-2}} \int\dd^dx\,\sqrt{-g}\,R.
\end{align}
For reference, the same action may be written in differential-form notation as
\begin{align}
  S_\text{H} = - \cfrac{2}{\kappa^{d-2}} \int \cfrac{1}{(d-2)!} \epsilon_{a_1\cdots a_d} e^{a_1}\wedge\cdots\wedge e^{a_{d-2}} \wedge R^{a_{d-1}a_d}.
\end{align}
This formula is included only to connect the component conventions used in the main text with the standard differential-form notation.

Appendix~\ref{app:density-spin connection} develops the corresponding identities for the density vierbein variables and the density spin connection. Unless stated otherwise, the paper uses the torsion-free second-order formulation, so the vierbein determines the spin connection.

\section{Density Spin Connection Identities}
\label{app:density-spin connection}

This appendix collects the identities involving the density vierbein variables and the density spin connection that are used in the main text. Appendix~\ref{app:vierbein-conventions} fixes the ordinary vierbein, torsion, spin connection, and curvature conventions. Here, we translate those conventions into density variables, assuming $d>2$ and $e>0$.

The density vierbein variables are defined by
\begin{align}\label{eq:appB_density_vierbein_definition}
  (E^\mu)_a \overset{\text{def}}{=} \sqrt e\,(e^\mu)_a.
\end{align}
Their inverse is defined by
\begin{align}\label{eq:appB_inverse_density_vierbein}
  (E_\mu)^a(E^\mu)_b &= \delta^a_b, & (E_\mu)^a(E^\nu)_a &= \delta^\nu_\mu.
\end{align}
In terms of the ordinary vierbein, one has
\begin{align}\label{eq:appB_inverse_density_vierbein_ordinary}
  (E_\mu)^a = \frac{1}{\sqrt e}(e_\mu)^a.
\end{align}

We use the following index conventions for density vierbein variables. Spacetime indices are raised and lowered with $g_{\mu\nu}$ and $g^{\mu\nu}$, while local Lorentz indices are raised and lowered with $\eta_{ab}$ and $\eta^{ab}$. The inverse density vierbein variables are recognized from the inverse relations~\eqref{eq:appB_inverse_density_vierbein}, not merely from the vertical position of the spacetime index. For example,
\begin{align}\label{eq:appB_density_vierbein_index_examples}
  (E^\mu)^a &\overset{\text{note}}{=} \eta^{ab}(E^\mu)_b, & (E_\mu)_a &\overset{\text{note}}{=} \eta_{ab}(E_\mu)^b, & (E_\alpha)^c(E^\beta)_c &= \delta^\beta_\alpha .
\end{align}
In particular,
\begin{align}
  g_{\mu\nu}(E^\nu)_a=e(E_\mu)_a.
\end{align}
Thus, lowering a spacetime index does not produce the inverse density vierbein; their density weights are opposite.

We denote the density vierbein determinant as $E = \det(E^\mu)_a$. Using $(E^\mu)_a=\sqrt e\,(e^\mu)_a$ and $\det(e^\mu)_a=e^{-1}$, one obtains
\begin{align}\label{eq:appB_density_determinant_relation}
  E=e^{(d-2)/2},\qquad e=E^{2/(d-2)}.
\end{align}
Consequently,
\begin{align}\label{eq:appB_density_logarithmic_identities}
  \begin{split}
    \pd_\mu\ln E &=(E_\sigma)^s\pd_\mu(E^\sigma)_s, \\
    \pd_\mu\ln e &=\frac{2}{d-2}(E_\sigma)^s\pd_\mu(E^\sigma)_s, \\
    \pd_\mu\ln\sqrt e &=\frac{1}{d-2}(E_\sigma)^s\pd_\mu(E^\sigma)_s .
  \end{split}
\end{align}

Differentiating the inverse relation $(E_\alpha)^a(E^\beta)_a=\delta^\beta_\alpha$ gives the inverse-matrix identity
\begin{align}\label{eq:appB_inverse_density_derivative}
  \pd_\mu(E_\alpha)^a = -(E_\beta)^a(E_\alpha)^b\pd_\mu(E^\beta)_b.
\end{align}
We use this identity repeatedly when we eliminate derivatives of inverse density vierbein variables.

We now derive the density spin connection. Starting from the ordinary tetrad postulate for the inverse vierbein,
\begin{align}\label{eq:appB_inverse_tetrad_postulate}
  \pd_\mu(e^\alpha)_a  + (\Gamma_\mu)^\alpha{}_\beta(e^\beta)_a - (e^\alpha)_b(\omega_\mu)^b{}_a =0,
\end{align}
we multiply by $\sqrt e$ and use
\begin{align}\label{eq:appB_density_vierbein_derivative}
  \pd_\mu(E^\alpha)_a = \sqrt e\,\pd_\mu(e^\alpha)_a + (\pd_\mu\ln\sqrt e)(E^\alpha)_a.
\end{align}
For a vector density of weight $1/2$, the density contribution is $-\tfrac12\Gamma^\rho{}_{\rho\mu}=-\partial_\mu\ln\sqrt e$. Hence the unabsorbed tetrad postulate is
\begin{align}\label{eq:appB_density_postulate_unabsorbed}
  0=\partial_\mu(E^\alpha)_a+(\Gamma_\mu)^\alpha{}_\beta(E^\beta)_a -(\partial_\mu\ln\sqrt e)(E^\alpha)_a-(E^\alpha)_b(\omega_\mu)^b{}_a.
\end{align}
Absorbing the density term gives $\nabla_\mu(E^\alpha)_a=0$ with the convention
\begin{align}\label{eq:appB_density_covariant_derivative}
  \nabla_\mu(E^\alpha)_a = \pd_\mu(E^\alpha)_a + (\Gamma_\mu)^\alpha{}_\beta(E^\beta)_a - (E^\alpha)_b(\Omega_\mu)^b{}_a .
\end{align}

The density spin connection is therefore
\begin{align}\label{eq:appB_density_spin_connection_definition}
  (\Omega_\mu)^a{}_b \overset{\text{def}}{=} (\omega_\mu)^a{}_b  + \pd_\mu\ln\sqrt e\,\delta^a_b.
\end{align}
Equivalently, using~\eqref{eq:appB_density_logarithmic_identities},
\begin{align}\label{eq:appB_density_spin_connection_definition_density}
  (\Omega_\mu)^a{}_b = (\omega_\mu)^a{}_b + \frac{1}{d-2}(E_\sigma)^s\pd_\mu(E^\sigma)_s\,\delta^a_b.
\end{align}
In the second-order formulation, $(\Omega_\mu)^a{}_b$ is not independent; it is the ordinary spin connection shifted by the trace term generated by the density weight of $(E^\mu)_a$.

Because the ordinary spin connection is antisymmetric,
\begin{align}\label{eq:appB_ordinary_spin_connection_antisymmetry}
  (\omega_\mu)_{ab}=-(\omega_\mu)_{ba},
\end{align}
the determinant of the density vierbein variables fixes the symmetric part of the density spin connection:
\begin{align}\label{eq:appB_density_spin_connection_symmetric_part}
  (\Omega_\mu)_{ab}+(\Omega_\mu)_{ba}  = \frac{2}{d-2}(E_\sigma)^s\pd_\mu(E^\sigma)_s\eta_{ab}.
\end{align}
Its trace is
\begin{align}\label{eq:appB_density_spin_connection_trace}
  (\Omega_\mu)^a{}_a  =  \frac{d}{d-2}(E_\sigma)^s\pd_\mu(E^\sigma)_s.
\end{align}
Thus, the density spin connection is not an ordinary Lorentz-algebra-valued connection. It acts on an upper Lorentz vector density of weight $-1/2$ as $\nabla_\mu V^a=\partial_\mu V^a+(\Omega_\mu)^a{}_bV^b$.

The antisymmetric part is unchanged by the trace shift:
\begin{align}\label{eq:appB_density_spin_connection_antisymmetric_part}
  (\Omega_\mu)_{ab}-(\Omega_\mu)_{ba} = 2(\omega_\mu)_{ab}.
\end{align}
We use this identity in both the auxiliary field relation and the fermion-gravity coupling.

We also use density frame components. They are defined by
\begin{align}\label{eq:appB_density_frame_derivative}
  \partial_m(E^\alpha)_a  \overset{\text{note}}{=} (E^\sigma)_m\pd_\sigma(E^\alpha)_a,
\end{align}
and
\begin{align}\label{eq:appB_density_frame_spin_connection}
  (\Omega_m)_{ab}  \overset{\text{def}}{=} (E^\mu)_m(\Omega_\mu)_{ab}.
\end{align}
The symbol $\partial_m$ denotes the density frame component of the derivative; it is not a derivative with respect to a coordinate $x^m$.

The local expressions $E_a=(E^\mu)_a\partial_\mu$ are differential operators in a chosen coordinate chart, rather than ordinary vector fields. In that chart, define
\begin{align}\label{eq:appB_density_anholonomy}
  [E_a,E_b]&=\mathfrak C_{ab}{}^c E_c, & \mathfrak C_{ab}{}^c&=\big[(E^\mu)_a\partial_\mu(E^\nu)_b-(E^\mu)_b\partial_\mu(E^\nu)_a\big](E_\nu)^c.
\end{align}
The relation to the ordinary frame coefficients is
\begin{align}
  \mathfrak C_{ab}{}^c=\sqrt e\,C_{ab}{}^c +\partial_a\ln\sqrt e\,\delta_b^c-\partial_b\ln\sqrt e\,\delta_a^c.
\end{align}
With $T_m=(E_\sigma)^s\partial_m(E^\sigma)_s$, the density spin connection takes the form
\begin{align}\label{eq:appB_density_connection_anholonomy}
  (\Omega_m)_{ab}=\frac12\big(\mathfrak C_{abm}+\mathfrak C_{mba}-\mathfrak C_{mab}\big) +\frac1{d-2}\big(T_m\eta_{ab}+T_b\eta_{ma}-T_a\eta_{mb}\big).
\end{align}
Equivalently, the inverse-frame derivative coefficients
\begin{align}
  \mathfrak D_{abc}&\overset{\text{def}}{=}(E^\mu)_a(E^\nu)_b\partial_\mu(E_\nu)_c,&
  \mathfrak C_{abc}&=\mathfrak D_{bac}-\mathfrak D_{abc}
\end{align}
provide an alternative way to obtain the same connection.

For compactness, Section~\ref{sec:Hilbert_action_perturbation} also uses
\begin{align}\label{eq:appB_X_definition}
  X_{ijk}  \overset{\text{def}}{=} (E_\alpha)_j\partial_i(E^\alpha)_k.
\end{align}
The slot convention implies $\mathfrak D_{abc}=-X_{acb}$. Its useful trace is
\begin{align}\label{eq:appB_X_trace}
  X_{ij}{}^j  =  (E_\alpha)^j\partial_i(E^\alpha)_j.
\end{align}
In particular, using~\eqref{eq:appB_density_frame_derivative} and~\eqref{eq:appB_density_logarithmic_identities},
\begin{align}\label{eq:appB_X_trace_determinant}
  X_{ij}{}^j  = (E^\mu)_i\pd_\mu\ln E.
\end{align}

The explicit expression for $(\Omega_m)_{ab}$ follows by substituting
\begin{align}\label{eq:appB_ordinary_density_substitution}
  (e^\mu)_a &= \frac{1}{\sqrt e}(E^\mu)_a, & (e_\mu)^a &= \sqrt e\,(E_\mu)^a 
\end{align}
into the torsion-free ordinary spin connection formula~\eqref{eq:appendix_spin_connection_anholonomy} and then using the determinant identities above. One obtains
\begin{align}\label{eq:appB_density_spin_connection_explicit}
  \begin{split}
    (\Omega_m)_{ab} = \Bigg[ &\frac12\left[(E^\sigma)_a\delta_b{}^c-(E^\sigma)_b\delta_a{}^c\right](E_\alpha)_m +\frac12\left[(E^\sigma)_m\delta_b{}^c-(E^\sigma)_b\delta_m{}^c\right](E_\alpha)_a \\
      &-\frac12\left[(E^\sigma)_m\delta_a{}^c-(E^\sigma)_a\delta_m{}^c\right](E_\alpha)_b \\
      &+\frac{1}{d-2}   \left[ (E^\sigma)_m\eta_{ab} +(E^\sigma)_b\eta_{ma} -(E^\sigma)_a\eta_{mb} \right](E_\alpha)^c  \Bigg]\pd_\sigma(E^\alpha)_c .
  \end{split}
\end{align}
This is the explicit density spin connection in terms of the density vierbein variables.

The formula~\eqref{eq:appB_density_spin_connection_explicit} reproduces the compatibility condition. Adding the same expression with $a$ and $b$ interchanged gives
\begin{align}\label{eq:appB_density_spin_connection_density_frame_compatibility}
  (\Omega_m)_{ab}+(\Omega_m)_{ba} = \frac{2}{d-2}(E^\mu)_m(E_\sigma)^s\pd_\mu(E^\sigma)_s\eta_{ab},
\end{align}
which is the density frame version of~\eqref{eq:appB_density_spin_connection_symmetric_part}.

Equation~\eqref{eq:appB_density_spin_connection_explicit} shows that the density spin connection is linear in the first derivatives of the density vierbein variables. However, its coefficients contain the inverse density vierbein variables. Therefore, it is non-polynomial in perturbation theory. For
\begin{align}\label{eq:appB_density_vierbein_perturbation}
  (E^\mu)_a=\delta^\mu_a+\kappa(\theta^\mu)_a,
\end{align}
the inverse density vierbein variables expand as
\begin{align}\label{eq:appB_inverse_density_vierbein_perturbation}
  (E_\mu)^a={}&\delta_\mu^a-\kappa\,\delta_\mu^b\delta_\nu^a(\theta^\nu)_b +\kappa^2\delta_\mu^b\delta_\rho^c\delta_\nu^a(\theta^\rho)_b(\theta^\nu)_c+O(\kappa^3).
\end{align}
Substitution into~\eqref{eq:appB_density_spin_connection_explicit} gives an infinite perturbative series.

The trace shift in the density spin connection does not change the curvature contribution entering the Hilbert action. Indeed,
\begin{align}\label{eq:appB_curvature_shift}
  R_{\mu\nu}{}^a{}_b[\Omega] = R_{\mu\nu}{}^a{}_b[\omega] + \left( \pd_\mu\pd_\nu\ln\sqrt e  -  \pd_\nu\pd_\mu\ln\sqrt e \right)\delta^a_b .
\end{align}
The last term vanishes locally for a smooth nonzero determinant. The commutator on a Lorentz vector density therefore gives the ordinary curvature, with
\begin{align}\label{eq:appB_curvature_conversion}
  R_{\mu\nu}{}^a{}_b[\Omega]=(E_\alpha)^a(E^\beta)_bR_{\mu\nu}{}^\alpha{}_\beta.
\end{align}
The usual all-local curvature components use the ordinary inverse vierbein on their first two indices:
\begin{align}\label{eq:appB_local_curvature_normalization}
  \begin{split}
    R_{mn}{}^a{}_b  &=(e^\mu)_m(e^\nu)_n R_{\mu\nu}{}^a{}_b =E^{-2/(d-2)}(E^\mu)_m(E^\nu)_n R_{\mu\nu}{}^a{}_b,\\
    R_{mn}&=R_{sn}{}^s{}_m,\qquad R=\eta^{mn}R_{mn}.
  \end{split}
\end{align}
The determinant factor is essential: conversion of both curvature two-form indices with density vierbeins alone would instead produce $eR_{mn}{}^a{}_b$.

The trace shift also drops out of antisymmetric contractions. In particular,
\begin{align}\label{eq:appB_antisymmetric_contraction_identity}
  &\left[ (E^\mu)_a(E^\nu)_b - (E^\mu)_b(E^\nu)_a \right] (\Omega_\nu)^{ab} = \left[ (E^\mu)_a(E^\nu)_b - (E^\mu)_b(E^\nu)_a \right] (\omega_\nu)^{ab}.
\end{align}
This is an algebraic consequence of the fact that the trace part of $(\Omega_\nu)^{ab}$ is proportional to $\eta^{ab}$.

The auxiliary field relation used in Section~\ref{sec:Hilbert_action_perturbation} is obtained by comparing the integrated-by-parts Hilbert action written with $\Omega$ to the auxiliary field coupling $A^i{}_{\alpha}{}^a\partial_i(E^\alpha)_a$. The result is
\begin{align}\label{eq:appB_auxiliary_field_density_spin_connection_relation}
  A^i{}_{\alpha}{}^a[\Omega,E] = \frac{2}{\kappa^{d-2}} \left[ (E_\alpha)^i \left( (\Omega_c)^{ac}-(\Omega_c)^{ca} \right) + (E_\alpha)^m \left( (\Omega_m)^{ia}-(\Omega_m)^{ai} \right) \right].
\end{align}
Substituting the metric-compatible density spin connection gives
\begin{align}\label{eq:appB_auxiliary_field_O_relation}
  A^i{}_{\alpha}{}^a[\Omega[E],E]  = \OO^i{}_{\alpha}{}^a{}^j{}_{\beta}{}^b\partial_j(E^\beta)_b.
\end{align}
This verifies the consistency between the geometric density spin-connection form and the auxiliary-field formulation.

Finally, the antisymmetric part of the density spin connection is related to the spin connection density that appears in the Dirac sector. Define
\begin{align}\label{eq:appB_Xi_definition}
  \Xi_{mab}  \overset{\text{def}}{=} e\,(\omega_m)_{ab}.
\end{align}
Since
\begin{align}\label{eq:appB_Omega_Xi_intermediate}
  (\Omega_m)_{ab}-(\Omega_m)_{ba}  = 2\sqrt e\,(\omega_m)_{ab},
\end{align}
one obtains
\begin{align}\label{eq:appB_Xi_Omega_relation}
  \Xi_{mab} = \frac{\sqrt e}{2} \left[ (\Omega_m)_{ab}-(\Omega_m)_{ba} \right].
\end{align}
This identity relates the geometric density spin connection to the connection density used in the Dirac action.

\section{Quadratic Kernels and Interaction Coefficients}
\label{app:feynman-rules}

This appendix collects the background propagators and momentum-space interaction coefficients of the corrected constrained polynomial constructed in Sections~\ref{sec:Hilbert_action_perturbation} and~\ref{sec:BRST}. The auxiliary propagator inverts the quadratic kernel on the background image. The interaction coefficients are derivatives of the specified polynomial representative; using them as a complete diagrammatic prescription additionally requires implementing the nonlinear restriction~\eqref{eq:shifted_auxiliary_image_constraint}. The functional measure and that independent-field implementation are outside the present classical construction. Throughout this appendix, we work around the flat density vierbein background and use the convenient gauge
\begin{align}\label{eq:appC_flat_gauge_choice}
  (\overline E^\mu)_a&=\delta^\mu_a, & \overline\Gamma&=0, & \overline\omega&=0, & \epsilon&=\frac12, & \rho&=1.
\end{align}

The fields in momentum space are defined by
\begin{align}\label{eq:appC_fourier_convention}
  \Phi(x) = \int \frac{\dd^d p}{(2\pi)^d}~ e^{\ii p\cdot x}\Phi(p),
\end{align}
so that a derivative acting on a field with momentum $p$ gives
\begin{align}\label{eq:appC_derivative_rule}
  \partial_\mu\Phi(x) \longrightarrow \ii p_\mu\Phi(p).
\end{align}
All momenta entering a vertex are taken as incoming, and every $n$-point vertex contains the momentum conservation factor
\begin{align}\label{eq:appC_momentum_conservation}
  (2\pi)^d\delta^{(d)}\!\left(\sum_{i=1}^{n}p_i\right).
\end{align}
In the formulae below, we suppress this overall factor.

The fields entering the rules are
\begin{align}\label{eq:appC_field_list}
  (\theta^\alpha)_a(p),\qquad B^i{}_{\alpha}{}^a(p),\qquad c^\mu(p),\qquad \overline c_\mu(p),\qquad c_{mn}(p),\qquad \overline\chi_{mn}(p).
\end{align}
The local Lorentz ghost and antighost are antisymmetric,
\begin{align}\label{eq:appC_lorentz_ghost_antisymmetry}
  c_{mn} &=-c_{nm}, & \overline\chi_{mn} &=-\overline\chi_{nm}.
\end{align}
The one-index ghost $c_m$ is the lowered diffeomorphism ghost, whereas $c_{mn}$ is the redefined local Lorentz ghost introduced in \eqref{eq:local_lorentz_ghost_redefinition_section4}.

The propagators are the inverses of the quadratic kernels in~\eqref{eq:quadratic_gauge_fixed_action_section4}, with the background auxiliary restriction understood. The momentum-space coefficients are defined by ordinary differentiation of the corrected polynomial representative. This differentiation specifies their tensor and momentum conventions and does not eliminate the nonlinear constraint. We retain the factor $\ii$ in the coefficient convention:
\begin{align}\label{eq:appC_vertex_definition}
  V_{\Phi_1\cdots\Phi_n}(p_1,\ldots,p_n) \overset{\text{def}}{=} \ii\,\frac{\delta^n S_\text{int}}{\delta\Phi_1(p_1)\cdots\delta\Phi_n(p_n)}.
\end{align}
For Grassmann-odd fields, we use left functional derivatives, applying the antighost derivative before the ghost derivative. Thus for an ordered term $\overline u\,M[\theta]v$, the vertex is $\ii M$ after the bosonic differentiation. Ghost lines are oriented from antighost to ghost, there is no permutation between these distinct legs, and each closed ghost loop contributes a factor of $-1$. With this convention, no extra symmetry factor is attached to an individual vertex after all functional derivatives have been taken. Diagrammatic symmetry factors are the usual ones associated with the perturbative expansion of a given graph. Momentum factors are generated only by the derivatives in the interaction action through the rule \eqref{eq:appC_derivative_rule}.

The Fourier convention fixes the derivative substitution
\begin{align}\label{eq:appC_propagating_denominator}
  \partial_\mu\partial^\mu\longrightarrow -p^2.
\end{align}
The signs of propagator denominators follow from the full quadratic kernels: the density vierbein pole has $p^2+\ii0$, whereas the diffeomorphism ghost kernel gives $-p^2+\ii0$. Denoting the complete density vierbein propagator by $(\mathcal G_\theta)^\alpha{}_a{}^\beta{}_b(p)$, we obtain
\begin{align}\label{eq:appC_theta_propagator}
  \left\langle 0 \middle| \mathrm{T}\!\left\{(\theta^\alpha)_a(p)(\theta^\beta)_b(-p)\right\}\middle|0\right\rangle = \ii\,\kappa^{d-4}\Bigg[\frac{\eta^{\alpha\beta}\eta_{ab} + \delta^\alpha{}_b\delta^\beta{}_a - \delta^\alpha{}_a\delta^\beta{}_b}{8(p^2+\ii0)}+\frac18\left(\eta^{\alpha\beta}\eta_{ab}-\delta^\alpha{}_b\delta^\beta{}_a\right)\Bigg].
\end{align}
The auxiliary field propagator is restricted to $\Img\OO_0$, is algebraic, and has no momentum pole:
\begin{align}\label{eq:appC_B_propagator}
  \left\langle 0 \middle| \mathrm{T}\!\left\{ B^{i_1}{}_{\alpha_1}{}^{a_1}(p)B^{i_2}{}_{\alpha_2}{}^{a_2}(-p) \right\}\middle|0\right\rangle = -\ii\,\kappa^{2-d} \mathscr{P}^{i_1}{}_{\alpha_1}{}^{a_1}{}^{i_2}{}_{\alpha_2}{}^{a_2}.
\end{align}
The diffeomorphism ghost propagator is
\begin{align}\label{eq:appC_diff_ghost_propagator}
  \left\langle 0 \middle| \mathrm{T}\!\left\{ c^\mu(p)\overline c_\nu(-p) \right\}\middle|0\right\rangle = \frac{\ii\,\kappa^{d-4}}{-p^2+\ii0}\,\delta^\mu_\nu .
\end{align}
Finally, the local Lorentz ghost propagator is
\begin{align}\label{eq:appC_lorentz_ghost_propagator}
  \left\langle 0 \middle| \mathrm{T}\!\left\{ c_{mn}(p)\overline\chi_{rs}(-p) \right\}\middle|0\right\rangle = \frac{\ii}{4}\kappa^{d-4} \left( \eta_{mr}\eta_{ns} - \eta_{ms}\eta_{nr} \right).
\end{align}
The normalization in \eqref{eq:appC_lorentz_ghost_propagator} assumes full summation over antisymmetric index pairs. Thus only $(\theta^\alpha)_a$ and the diffeomorphism ghost carry momentum poles. The auxiliary field $B$ and the local Lorentz ghost sector are algebraic in this gauge.

The diagrams below denote the background quadratic contractions:
\begin{align}
  \begin{gathered}
    \begin{fmffile}{D01_propagator_theta}
      \begin{fmfgraph*}(40,40)
        \fmfleft{L}
        \fmfright{R}
        \fmf{dbl_wiggly,label=$p$}{L,R}
        \fmflabel{${}^\alpha_a$}{L}
        \fmflabel{${}^\beta_b$}{R}
      \end{fmfgraph*}
    \end{fmffile}
  \end{gathered}
  \hspace{20pt} &= (\mathcal G_\theta)^\alpha{}_a{}^\beta{}_b(p) ,\\
  \begin{gathered}
    \begin{fmffile}{D01_propagator_B}
      \begin{fmfgraph*}(40,40)
        \fmfleft{L}
        \fmfright{R}
        \fmf{dbl_dashes}{L,R}
        \fmflabel{${}^{i_1}{}_{\alpha_1}{}^{a_1}$}{L}
        \fmflabel{${}^{i_2}{}_{\alpha_2}{}^{a_2}$}{R}
      \end{fmfgraph*}
    \end{fmffile}
  \end{gathered}
  \hspace{30pt}
  &= -\ii\,\kappa^{2-d}\,
  \mathscr{P}^{i_1}{}_{\alpha_1}{}^{a_1}{}^{i_2}{}_{\alpha_2}{}^{a_2}, \\
  \begin{gathered}
    \begin{fmffile}{D01_propagator_c}
      \begin{fmfgraph*}(40,40)
        \fmfleft{L}
        \fmfright{R}
        \fmf{dots,label=$p$}{L,R}
        \fmflabel{${}^\mu$}{L}
        \fmflabel{${}_\nu$}{R}
      \end{fmfgraph*}
    \end{fmffile}
  \end{gathered} \hspace{10pt}&= \ii \, \kappa^{d-4} \, \cfrac{1}{-p^2 + \ii 0} \, \delta^\mu_\nu , \\
  \begin{gathered}
    \begin{fmffile}{D01_propagator_cc}
      \begin{fmfgraph*}(40,40)
        \fmfleft{L}
        \fmfright{R}
        \fmf{dbl_dots,label=$p$}{L,R}
        \fmflabel{${}_{mn}$}{L}
        \fmflabel{${}_{rs}$}{R}
      \end{fmfgraph*}
    \end{fmffile}
  \end{gathered} \hspace{15pt} & = \ii\, \kappa^{d-4} \, \cfrac14 \left( \eta_{mr}\eta_{ns} - \eta_{ms}\eta_{nr} \right) .
\end{align}

We now record the six interaction coefficient tensors of the corrected constrained polynomial. The basic background tensors are $\mathscr P$ and $\mathscr P^{-1}$, defined in~\eqref{eq:background_P_operator_section3} and~\eqref{eq:background_P_inverse_section3}. The pure quartic tensor retains the expression
\begin{align}\label{eq:appC_V4_definition}
  \begin{split}
    (\mathcal{V}_{\theta\theta\theta\theta})^{j_1}{}_{\beta_1}{}^{b_1}{}_{\gamma_1}{}^{c_1}{}_{\gamma_2}{}^{c_2}{}^{j_2}{}_{\beta_2}{}^{b_2} \overset{\text{def}}{=}  &\frac18\eta_{i_1i_2}\mathscr{P}^{i_1}{}_{\gamma_1}{}^{c_2}{}^{j_1}{}_{\beta_1}{}^{b_1} \mathscr{P}^{i_2}{}_{\gamma_2}{}^{c_1}{}^{j_2}{}_{\beta_2}{}^{b_2} -\frac{1}{16}\eta_{i_1i_2}\mathscr{P}^{i_1}{}_{\gamma_1}{}^{c_1}{}^{j_1}{}_{\beta_1}{}^{b_1}
  \mathscr{P}^{i_2}{}_{\gamma_2}{}^{c_2}{}^{j_2}{}_{\beta_2}{}^{b_2}.
  \end{split}
\end{align}
The coefficients are most directly specified by the primitive matrices $P,q$ and the elementary middle-index and derivative-index maps.

The pure density vierbein interactions are generated by
\begin{align}\label{eq:appC_pure_theta_interactions}
  \begin{split}
    S_{\theta^3}+S_{\theta^4}= \int\dd^dx\Big[& \kappa^{5-d}\,\partial_{j_1}(\theta^{\beta_1})_{b_1} (\mathcal{V}_{\theta\theta\theta})^{j_1}{}_{\beta_1}{}^{b_1}{}_{\gamma}{}^{c}{}^{j_2}{}_{\beta_2}{}^{b_2} (\theta^\gamma)_c\partial_{j_2}(\theta^{\beta_2})_{b_2}\\
      &-\kappa^{6-d}\,\partial_{j_1}(\theta^{\beta_1})_{b_1} (\mathcal{V}_{\theta\theta\theta\theta})^{j_1}{}_{\beta_1}{}^{b_1}{}_{\gamma_1}{}^{c_1}{}_{\gamma_2}{}^{c_2}{}^{j_2}{}_{\beta_2}{}^{b_2} (\theta^{\gamma_1})_{c_1}(\theta^{\gamma_2})_{c_2}\partial_{j_2}(\theta^{\beta_2})_{b_2} \Big].
  \end{split}
\end{align}
The derivatives in \eqref{eq:appC_pure_theta_interactions} determine which external legs carry momentum factors.

The interactions involving the auxiliary field are generated by
\begin{align}\label{eq:appC_B_interactions}
  \begin{split}
    S_{B,\text{int}}=\int\dd^dx\Big[& -\kappa^{d-1}\,B^{i_1}{}_{\alpha_1}{}^{a_1} (\mathcal{V}_{\theta BB})_{i_1}{}^{\alpha_1}{}_{a_1}{}_{i_2}{}^{\alpha_2}{}_{a_2}{}_{\gamma}{}^c (\theta^\gamma)_c B^{i_2}{}_{\alpha_2}{}^{a_2}\\
      &-\kappa^d\,B^{i_1}{}_{\alpha_1}{}^{a_1} (\mathcal{V}_{\theta\theta BB})_{i_1}{}^{\alpha_1}{}_{a_1}{}_{i_2}{}^{\alpha_2}{}_{a_2}{}_{\gamma_1}{}^{c_1}{}_{\gamma_2}{}^{c_2} (\theta^{\gamma_1})_{c_1}(\theta^{\gamma_2})_{c_2}B^{i_2}{}_{\alpha_2}{}^{a_2}\\
      &+\kappa^2\,B^{i_1}{}_{\alpha_1}{}^{a_1} (\mathcal{V}_{\theta\theta B})_{i_1}{}^{\alpha_1}{}_{a_1}{}_{\gamma}{}^c{}^j{}_{\beta}{}^b (\theta^\gamma)_c\partial_j(\theta^\beta)_b\\
      &-\kappa^3\,B^{i_1}{}_{\alpha_1}{}^{a_1} (\mathcal{V}_{\theta\theta\theta B})_{i_1}{}^{\alpha_1}{}_{a_1}{}_{\gamma_1}{}^{c_1}{}_{\gamma_2}{}^{c_2}{}^j{}_{\beta}{}^b (\theta^{\gamma_1})_{c_1}(\theta^{\gamma_2})_{c_2}\partial_j(\theta^\beta)_b \Big].
  \end{split}
\end{align}
Thus the finite auxiliary field vertex classes are
\begin{align}\label{eq:appC_B_vertex_classes}
  \theta BB,\qquad \theta^2BB,\qquad \theta^2B,
  \qquad \theta^3B.
\end{align}
No derivatives act on the $B$ legs.

The corresponding diagrams are given below:
\begin{align}
  \nonumber \\
  \begin{gathered}
    \begin{fmffile}{V_theta_theta_theta}
      \begin{fmfgraph*}(50,70)
        \fmftop{T}
        \fmfleft{L}
        \fmfright{R}
        \fmf{dbl_wiggly,label=$p_1$}{L,V}
        \fmf{dbl_wiggly,label=$p_2$}{V,T}
        \fmf{dbl_wiggly,label=$p_3$}{V,R}
        \fmfdot{V}
        \fmflabel{${}^{\alpha_1}_{a_1}$}{L}
        \fmflabel{${}^{\alpha_2}_{a_2}$}{T}
        \fmflabel{${}^{\alpha_3}_{a_3}$}{R}
      \end{fmfgraph*}
    \end{fmffile}
  \end{gathered}
  \hspace{20pt}
  \begin{split}
    = -\ii\,\kappa^{5-d} \Big[ & (p_1)_{j_1} (p_3)_{j_2} (\mathcal{V}_{\theta\theta\theta})^{j_1}{}_{\alpha_1}{}^{a_1}{}_{\alpha_2}{}^{a_2}{}^{j_2}{}_{\alpha_3}{}^{a_3} \\
      & + \text{all permutations of }  (p_1,\alpha_1,a_1),\,  (p_2,\alpha_2,a_2),\,   (p_3,\alpha_3,a_3) \Big],
  \end{split}
\end{align}
\begin{align}
  \begin{gathered}
    \begin{fmffile}{V_theta_theta_theta_theta}
      \begin{fmfgraph*}(60,60)
        \fmftop{T}
        \fmfbottom{B}
        \fmfleft{L}
        \fmfright{R}
        \fmf{dbl_wiggly,label=$p_1$}{L,V}
        \fmf{dbl_wiggly,label=$p_2$}{V,T}
        \fmf{dbl_wiggly,label=$p_3$}{V,R}
        \fmf{dbl_wiggly,label=$p_4$}{V,B}
        \fmfdot{V}
        \fmflabel{${}^{\alpha_1}_{a_1}$}{L}
        \fmflabel{${}^{\alpha_2}_{a_2}$}{T}
        \fmflabel{${}^{\alpha_3}_{a_3}$}{R}
        \fmflabel{${}^{\alpha_4}_{a_4}$}{B}
      \end{fmfgraph*}
    \end{fmffile}
  \end{gathered}
  \hspace{20pt}=
  \begin{split}
    +\ii\,\kappa^{6-d} \Big[ & (p_1)_{j_1} (p_4)_{j_2} (\mathcal{V}_{\theta\theta\theta\theta})^{j_1}{}_{\alpha_1}{}^{a_1}{}_{\alpha_2}{}^{a_2}{}_{\alpha_3}{}^{a_3}{}^{j_2}{}_{\alpha_4}{}^{a_4} \\
      &+\text{all permutations of } (p_i,\alpha_i,a_i),\quad i=1,\ldots,4
  \Big],
  \end{split}
\end{align}

\begin{align}
  \begin{gathered}
    \begin{fmffile}{V_theta_B_B}
      \begin{fmfgraph*}(50,70)
        \fmftop{L}
        \fmfleft{T}
        \fmfright{R}
        \fmf{dbl_wiggly,label=$p_1$}{L,V}
        \fmf{dbl_dashes,label=$p_2$}{V,T}
        \fmf{dbl_dashes,label=$p_3$}{V,R}
        \fmfdot{V}
        \fmflabel{${}^{\alpha_1}_{a_1}$}{L}
        \fmflabel{${}^{i_2}{}_{\alpha_2}{}^{a_2}$}{T}
        \fmflabel{${}^{i_3}{}_{\alpha_3}{}^{a_3}$}{R}
      \end{fmfgraph*}
    \end{fmffile}
  \end{gathered}
  \hspace{30pt} =
  \begin{split}
    -\ii\,\kappa^{d-1} \Big[ &  (\mathcal{V}_{\theta BB})_{i_2}{}^{\alpha_2}{}_{a_2}{}_{i_3}{}^{\alpha_3}{}_{a_3}{}_{\alpha_1}{}^{a_1} \\
      & +\text{the permutation of $B$-legs } (p_2,i_2,\alpha_2,a_2) \leftrightarrow (p_3,i_3,\alpha_3,a_3)\Big],
  \end{split}
\end{align}

\begin{align}
  \begin{gathered}
    \begin{fmffile}{V_theta_theta_B_B}
      \begin{fmfgraph*}(60,60)
        \fmftop{T}
        \fmfbottom{B}
        \fmfleft{L}
        \fmfright{R}
        \fmf{dbl_wiggly,label=$p_1$}{L,V}
        \fmf{dbl_wiggly,label=$p_2$}{V,T}
        \fmf{dbl_dashes,label=$p_3$}{V,R}
        \fmf{dbl_dashes,label=$p_4$}{V,B}
        \fmfdot{V}
        \fmflabel{${}^{\alpha_1}_{a_1}$}{L}
        \fmflabel{${}^{\alpha_2}_{a_2}$}{T}
        \fmflabel{${}^{i_3}{}_{\alpha_3}{}^{a_3}$}{R}
        \fmflabel{${}^{i_4}{}_{\alpha_4}{}^{a_4}$}{B}
      \end{fmfgraph*}
    \end{fmffile}
  \end{gathered}
  \hspace{30pt} =
  \begin{split}
    -\ii\,\kappa^{d}\Big[& (\mathcal{V}_{\theta\theta BB})_{i_3}{}^{\alpha_3}{}_{a_3}{}_{i_4}{}^{\alpha_4}{}_{a_4}{}_{\alpha_1}{}^{a_1}{}_{\alpha_2}{}^{a_2} \\
      &+\text{all independent permutations of $\theta$-legs } (p_1,\alpha_1,a_1),(p_2,\alpha_2,a_2) \\
      & \text{and the two labelled $B$-legs }  (p_3,i_3,\alpha_3,a_3),(p_4,i_4,\alpha_4,a_4) \Big],
  \end{split}
\end{align}

\begin{align}
  \begin{gathered}
    \begin{fmffile}{V_theta_theta_B}
      \begin{fmfgraph*}(50,70)
        \fmftop{R}
        \fmfleft{L}
        \fmfright{T}
        \fmf{dbl_wiggly,label=$p_1$}{L,V}
        \fmf{dbl_wiggly,label=$p_2$}{V,T}
        \fmf{dbl_dashes,label=$p_3$}{V,R}
        \fmfdot{V}
        \fmflabel{${}^{\alpha_1}_{a_1}$}{L}
        \fmflabel{${}^{\alpha_2}_{a_2}$}{T}
        \fmflabel{${}^{i_3}{}_{\alpha_3}{}^{a_3}$}{R}
      \end{fmfgraph*}
    \end{fmffile}
  \end{gathered}
  \hspace{30pt} =
  \begin{split}
    -\,\kappa^{2} \Big[& (p_2)_{j} (\mathcal{V}_{\theta\theta B})_{i_3}{}^{\alpha_3}{}_{a_3}{}_{\alpha_1}{}^{a_1}{}^{j}{}_{\alpha_2}{}^{a_2} \\
      & +\text{the permutation of the two $\theta$-legs } (p_1,\alpha_1,a_1) \leftrightarrow(p_2,\alpha_2,a_2)\Big].
  \end{split}
\end{align}

\begin{align}
  \begin{gathered}
    \begin{fmffile}{V_theta_theta_theta_B}
      \begin{fmfgraph*}(60,60)
        \fmftop{T}
        \fmfbottom{B}
        \fmfleft{L}
        \fmfright{R}
        \fmf{dbl_wiggly,label=$p_1$}{L,V}
        \fmf{dbl_wiggly,label=$p_2$}{V,T}
        \fmf{dbl_wiggly,label=$p_3$}{V,R}
        \fmf{dbl_dashes,label=$p_4$}{V,B}
        \fmfdot{V}
        \fmflabel{${}^{\alpha_1}_{a_1}$}{L}
        \fmflabel{${}^{\alpha_2}_{a_2}$}{T}
        \fmflabel{${}^{\alpha_3}_{a_3}$}{R}
        \fmflabel{${}^{i_4}{}_{\alpha_4}{}^{a_4}$}{B}
      \end{fmfgraph*}
    \end{fmffile}
  \end{gathered}
  \hspace{20pt} =
  \begin{split}
    +\,\kappa^{3}\Big[ & (p_3)_{j} (\mathcal{V}_{\theta\theta\theta B})_{i_4}{}^{\alpha_4}{}_{a_4}{}_{\alpha_1}{}^{a_1}{}_{\alpha_2}{}^{a_2}{}^{j}{}_{\alpha_3}{}^{a_3} \\
      & +\text{all permutations of the three $\theta$-legs } (p_1,\alpha_1,a_1),\, (p_2,\alpha_2,a_2),\, (p_3,\alpha_3,a_3)\Big]. 
  \end{split} 
\end{align}

{~}\\

The ghost interactions follow from \eqref{eq:ghost_interactions_section4}. They contain a single density vierbein perturbation, one antighost, and either the diffeomorphism ghost or the redefined local Lorentz ghost. First, it is useful to display the coordinate-space form and then extract the momentum-space rules. The diffeomorphism antighost part is
\begin{align}\label{eq:appC_diff_ghost_interactions}
  \begin{split}
    S_{\overline c,\text{int}} = \kappa^{5-d}  \int   \dd^dx &~(\partial_\nu\overline c_\mu) \Bigg[ \eta^{\nu a} \left( c^\rho\partial_\rho(\theta^\mu)_a - (\theta^\rho)_a\partial_\rho c^\mu + \frac12(\theta^\mu)_a\partial_\rho c^\rho + \left[c_a{}^b+\frac12(\partial_a c^b-\partial^b c_a)\right](\theta^\mu)_b \right) \\
      & + \eta^{\mu a} \left( c^\rho\partial_\rho(\theta^\nu)_a - (\theta^\rho)_a\partial_\rho c^\nu + \frac12(\theta^\nu)_a\partial_\rho c^\rho + \left[c_a{}^b+\frac12(\partial_a c^b-\partial^b c_a)\right](\theta^\nu)_b \right) \Bigg].
  \end{split}
\end{align}
The local Lorentz antighost part is
\begin{align}\label{eq:appC_lorentz_ghost_interactions}
  \begin{split}
    S_{\overline\chi,\text{int}}  =  -\kappa^{5-d} & \!\!\int\!\!\dd^dx  \, \overline\chi^{mn} \Bigg[ \eta_{m\mu} \left( c^\rho\partial_\rho(\theta^\mu)_n - (\theta^\rho)_n\partial_\rho c^\mu + \frac12(\theta^\mu)_n\partial_\rho c^\rho + \left[c_n{}^s+\frac12(\partial_n c^s-\partial^s c_n)\right](\theta^\mu)_s \right) \\
      & - \eta_{n\mu} \left( c^\rho\partial_\rho(\theta^\mu)_m - (\theta^\rho)_m\partial_\rho c^\mu +\frac12(\theta^\mu)_m\partial_\rho c^\rho  +  \left[c_m{}^s+\frac12(\partial_m c^s  - \partial^s c_m ) \right](\theta^\mu)_s \right)\Bigg].
  \end{split}
\end{align}
The resulting ghost vertex classes are
\begin{align}\label{eq:appC_ghost_vertex_classes}
  \overline c\,c\,\theta,
  \qquad
  \overline c\,c_{mn}\,\theta,
  \qquad
  \overline\chi\,c\,\theta,
  \qquad
  \overline\chi\,c_{mn}\,\theta.
\end{align}
Here $c_m$ denotes the lowered diffeomorphism ghost, while $c_{mn}$ denotes the redefined local Lorentz ghost. In labels such as $\mathcal{V}_{\overline c c_{mn}\theta}$ and $\mathcal{V}_{\overline\chi c_{mn}\theta}$, the symbol $c_{mn}$ denotes the two-index ghost species; the explicit indices of a particular external two-index ghost leg are displayed separately, for example as $rs$. In the flat-background gauge \eqref{eq:appC_flat_gauge_choice}, with the linear gauge-fixing functions used in Section~\ref{sec:BRST}, there are no higher ghost vertices because the BRST variation of $(\theta^\mu)_a$ is at most linear in $\theta$.

The ghost interaction tensors are defined as follows:
\begin{align}\label{eq:appC_W_cbar_c_dtheta}
  \begin{split}
    (\mathcal{W}_{\partial\overline c\,c\,\partial\theta})^{i\mu}{}_{\rho}{}^{j}{}_{\alpha}{}^{a} \overset{\text{def}}{=}&~ \eta^{ia}\delta^\mu{}_{\alpha}\delta^j{}_{\rho} + \eta^{\mu a}\delta^i{}_{\alpha}\delta^j{}_{\rho},  \\
    (\mathcal{W}_{\partial\overline c\,\theta\,\partial c})^{i\mu}{}_{\alpha}{}^{a}{}^{j}{}_{\rho} \overset{\text{def}}{=}&~ -\eta^{ia}\delta_\alpha{}^j\delta^\mu{}_{\rho} + \frac12\eta^{ia}\delta^\mu{}_{\alpha}\delta^j{}_{\rho} + \frac12\delta^\mu{}_{\alpha} \left( \eta^{ij}\delta^a{}_{\rho} - \delta^i{}_{\rho}\eta^{aj} \right)\\
    &- \eta^{\mu a}\delta_\alpha{}^j\delta^i{}_{\rho} + \frac12\eta^{\mu a}\delta^i{}_{\alpha}\delta^j{}_{\rho} + \frac12\delta^i{}_{\alpha} \left( \eta^{\mu j}\delta^a{}_{\rho} - \delta^\mu{}_{\rho}\eta^{aj} \right), \\
    (\mathcal{W}_{\partial\overline c\,c_2\,\theta})^{i\mu}{}^{rs}{}_{\alpha}{}^{a} \overset{\text{def}}{=}&~ \frac12\delta^\mu{}_{\alpha} \left( \eta^{ir}\eta^{as} - \eta^{is}\eta^{ar} \right) + \frac12\delta^i{}_{\alpha} \left( \eta^{\mu r}\eta^{as} - \eta^{\mu s}\eta^{ar} \right), \\
    (\mathcal{W}_{\overline\chi\,c\,\partial\theta})_{mn\rho}{}^{j}{}_{\alpha}{}^{a} \overset{\text{def}}{=}&~ - \left(\eta_{m\alpha}\delta_n{}^a -\eta_{n\alpha}\delta_m{}^a \right) \delta_\rho{}^j, \\
    (\mathcal{W}_{\overline\chi\,\theta\,\partial c})_{mn}{}_{\alpha}{}^{a}{}^{j}{}_{\rho} \overset{\text{def}}{=}&~ \eta_{m\rho}\delta_n{}^a\delta_\alpha{}^j - \eta_{n\rho}\delta_m{}^a\delta_\alpha{}^j -\frac12 \left( \eta_{m\alpha}\delta_n{}^a - \eta_{n\alpha}\delta_m{}^a \right) \delta_\rho{}^j \\
    &- \frac12\eta_{m\alpha}\left( \delta_n{}^j\delta^a{}_{\rho} - \eta^{aj}\eta_{n\rho} \right) + \frac12\eta_{n\alpha} \left( \delta_m{}^j\delta^a{}_{\rho} - \eta^{aj}\eta_{m\rho} \right), \\
    (\mathcal{W}_{\overline\chi\,c_2\,\theta})_{mn}{}^{rs}{}_{\alpha}{}^{a} \overset{\text{def}}{=}&~ -\frac12\eta_{m\alpha} \left( \delta_n{}^r\eta^{as} - \delta_n{}^s\eta^{ar} \right) + \frac12\eta_{n\alpha} \left( \delta_m{}^r\eta^{as} - \delta_m{}^s\eta^{ar} \right).
  \end{split}
\end{align}
The tensors with the index pair $rs$ are antisymmetric in $r,s$. The tensors in the local Lorentz antighost sector are antisymmetric in $m,n$. The label $c_2$ in the subscripts of the $\mathcal{W}$-tensors denotes the two-index ghost $c_{mn}$ and does not denote an additional field.

For the diagrams below, dotted lines denote the diffeomorphism ghost sector, double-dotted lines denote the redefined local Lorentz ghost sector, and double-wiggly lines denote the density vierbein perturbation. All momenta are incoming.
\begin{align}
  \begin{gathered}
    \begin{fmffile}{V_cbar_c_theta}
      \begin{fmfgraph*}(50,70)
        \fmftop{T}
        \fmfleft{L}
        \fmfright{R}
        \fmf{dots,label=$p_1$}{L,V}
        \fmf{dots,label=$p_2$}{V,R}
        \fmf{dbl_wiggly,label=$k$}{V,T}
        \fmfdot{V}
        \fmflabel{${}_{\mu}$}{L}
        \fmflabel{${}^{\rho}$}{R}
        \fmflabel{${}^{\alpha}_{a}$}{T}
      \end{fmfgraph*}
    \end{fmffile}
  \end{gathered}
  \hspace{20pt}
  &= - \ii\,\kappa^{5-d} \Big[ (p_1)_i (k)_j (\mathcal{W}_{\partial\overline c\,c\,\partial\theta})^{i\mu}{}_{\rho}{}^{j}{}_{\alpha}{}^{a} + (p_1)_i (p_2)_j (\mathcal{W}_{\partial\overline c\,\theta\,\partial c})^{i\mu}{}_{\alpha}{}^{a}{}^{j}{}_{\rho} \Big] ,
\end{align}
\begin{align}
  \begin{gathered}
    \begin{fmffile}{V_cbar_cmn_theta}
      \begin{fmfgraph*}(50,70)
        \fmftop{T}
        \fmfleft{L}
        \fmfright{R}
        \fmf{dots,label=$p$}{L,V}
        \fmf{dbl_dots,label=$q$}{V,R}
        \fmf{dbl_wiggly,label=$k$}{V,T}
        \fmfdot{V}
        \fmflabel{${}_{\mu}$}{L}
        \fmflabel{${}_{rs}$}{R}
        \fmflabel{${}^{\alpha}_{a}$}{T}
      \end{fmfgraph*}
    \end{fmffile}
  \end{gathered}
  \hspace{20pt}
  &= - \,\kappa^{5-d} (p)_i (\mathcal{W}_{\partial\overline c\,c_2\,\theta})^{i\mu}{}^{rs}{}_{\alpha}{}^{a} ,
\end{align}
\begin{align}
  \begin{gathered}
    \begin{fmffile}{V_chibar_c_theta}
      \begin{fmfgraph*}(50,70)
        \fmftop{T}
        \fmfleft{L}
        \fmfright{R}
        \fmf{dbl_dots,label=$q$}{L,V}
        \fmf{dots,label=$p$}{V,R}
        \fmf{dbl_wiggly,label=$k$}{V,T}
        \fmfdot{V}
        \fmflabel{${}^{mn}$}{L}
        \fmflabel{${}^{\rho}$}{R}
        \fmflabel{${}^{\alpha}_{a}$}{T}
      \end{fmfgraph*}
    \end{fmffile}
  \end{gathered}
  \hspace{20pt}
  &= -\,\kappa^{5-d} \Big[ (k)_j (\mathcal{W}_{\overline\chi\,c\,\partial\theta})_{mn\rho}{}^{j}{}_{\alpha}{}^{a} + (p)_j (\mathcal{W}_{\overline\chi\,\theta\,\partial c})_{mn}{}_{\alpha}{}^{a}{}^{j}{}_{\rho} \Big] ,
\end{align}
\begin{align}
  \begin{gathered}
    \begin{fmffile}{V_chibar_cmn_theta}
      \begin{fmfgraph*}(50,70)
        \fmftop{T}
        \fmfleft{L}
        \fmfright{R}
        \fmf{dbl_dots,label=$q_1$}{L,V}
        \fmf{dbl_dots,label=$q_2$}{V,R}
        \fmf{dbl_wiggly,label=$k$}{V,T}
        \fmfdot{V}
        \fmflabel{${}^{mn}$}{L}
        \fmflabel{${}_{rs}$}{R}
        \fmflabel{${}^{\alpha}_{a}$}{T}
      \end{fmfgraph*}
    \end{fmffile}
  \end{gathered}
  \hspace{20pt}
  &= \ii\,\kappa^{5-d} (\mathcal{W}_{\overline\chi\,c_2\,\theta})_{mn}{}^{rs}{}_{\alpha}{}^{a} .
\end{align}

All displayed momentum-space coefficients follow from the specified polynomial by symbolic differentiation and by applying the derivative-placement convention \eqref{eq:appC_derivative_rule}. This is the natural implementation strategy in computer-algebra tools. In particular, \texttt{FeynGrav} provides a framework for gravity Feynman rules within \texttt{FeynCalc}~\cite{Latosh:2022ydd}, and its later versions implement finite ghost-graviton structures and Cheung-Remmen variables~\cite{Latosh:2025vax}. These expressions specify the constrained polynomial; a complete diagrammatic implementation must additionally enforce its auxiliary restriction. The ghost sector is independent of this hierarchy and requires, in addition, the six $\mathcal{W}$-tensors defined in \eqref{eq:appC_W_cbar_c_dtheta}.

\bibliographystyle{unsrturl}
\bibliography{HFcVsG}

\end{document}